\documentclass[aps,preprint]{revtex4}%
\usepackage{amsfonts}
\usepackage{amsmath}
\usepackage{amssymb}
\usepackage{graphicx}%
\begin{document}
\title{Effect of spacetime dimensions on quantum entanglement between two uniformly
accelerated atoms}
\author{Jiatong Yan}
\author{Baocheng Zhang}
\email{zhangbaocheng@cug.edu.cn}
\affiliation{School of Mathematics and Physics, China University of Geosciences, Wuhan
430074, China}
\pacs{PACS number}

\begin{abstract}
We investigate the entanglement dynamics for a quantum system composed of two
uniformly accelerated Unruh-DeWitt detectors in different spacetime
dimensions. It is found that the range of parameters in which entanglement can
be generated is expanded but the amount of generated entanglement is decreased
with the increasing spacetime dimension, by calculating the evolution of
two-atom states using the method for open quantum systems. We study the
entanglement evolution between two accelerated atoms for different initial
two-atom states, and the influence of corresponding spacetime dimensions for
every initial state is discussed. When the spacetime dimensions increase, the
change of entanglement becomes slower with time. The influence of spacetime
dimensions on the change of entanglement also expands to the case of the
massive field. The time delay for entanglement generation is shown in
different spacetime dimensions. In particular, entanglement decreases more
quickly with the increasing spacetime dimensions compared with that in the
case of the massless field. The recent found anti-Unruh effect is discussed,
and a novel and interesting phenomenon is found that the Unruh effect in small
spacetime dimensions can become the anti-Unruh effect in large spacetime
dimensions with the same parameters.

\end{abstract}
\maketitle
\tableofcontents

\newpage

\section{Introduction}

When studying black hole evaporation, Unruh found that a uniformly
accelerating observer in the Minkowski vacuum with acceleration $a$ would
perceive a thermal bath of particles with the temperature $T={\hbar}{a}%
/{{2\pi}{c}{k_{B}}}$ \cite{wgu76,pd75,saf73}, which is dubbed as Unruh effect.
Since then, it has been extended to many different situations (see the review
\cite{chm08} and references therein). Some of them are related to the physical
effects such as the dynamic Casimir effect \cite{cp11,rt08,aas06}, the Lamb
shifts \cite{am95,rp98}, the resonance interactions \cite{rzp16,zpr16} and so on.

Another closely related physical phenomenon influenced by the Unruh effect is
quantum entanglement. For a maximally entangled bipartite quantum state, it
was found that the state would become less entangled \cite{fm05} and even
sudden death \cite{dss15} to an observer in relative acceleration. The
decrease of entanglement in these and other cases
\cite{amt06,ml09,mgl10,wj11,ses12,bfl12,ro15,hy15} could attribute to the fact
that accelerating observers only have partial access to the information
encoded in the quantum entanglement. However, several studies
\cite{rrs05,sm09,mm11} based on the famous Unruh-DeWitt detector model
\cite{dhi79} have found that quantum entanglement could be enhanced by\ the
Unruh effect when coupling one or two detectors into the local quantum fields
even if they were spacelike separated. This entanglement enhancement is
speculated to be extracted from the quantum entanglement of vacuum with which
the accelerated detectors interacted, by a mechanism called entanglement
harvesting \cite{br03,ctm19,zy20,gtm21,cm22}. But these enhancement phenomena
didn't represent a stationary mechanism.

A recent study called as the anti-Unruh effect \cite{bmm16} can provide a
stationary mechanism \cite{gmr16} for entanglement enhancement. The anti-Unruh
effect means that a uniformly accelerating particle detector may cool down in
certain conditions, opposite to the normal Unruh effect. It has been shown to
represent a general stationary mechanism that can exist under a stationary
state satisfying the Kubo-Martin-Schwinger (KMS) condition
\cite{rk57,ms59,fjl16} and is independent on any kind of boundary conditions
\cite{bmm16,gmr16}. A recent calculation showed that the anti-Unruh effect can
lead to an increase in the quantum entanglement for the bipartite
\cite{lzy18,zhy21,chy22} and many-body \cite{pz20} quantum states. The
experimental feasibility of testing the anti-Unruh effect was theoretically
analyzed using this multi-body state accelerated in the thermal environment
\cite{pz21,bm21}. Interestingly, the anti-Unruh effect can be applied to
Banados-Teitelboim-Zanelli (BTZ) black holes \cite{btz92} and presented a
novel phenomenon called the anti-Hawking phenomena \cite{hhz20,rhm21}.

It is noted that all these works listed above about the fascinating change of
entanglement induced by the acceleration were discussed in the $2$ or $4$
spacetime dimensions. But the response function of the accelerated detector in
the vacuum, which is crucial for calculating the change of entanglement, is
dependent on the number of spacetime dimensions \cite{st86,ls02}. In even
spacetime dimensions, an accelerating observer would feel a Bose-Einstein
distribution for the Bosonic field and a Fermi-Dirac distribution for the
Fermionic field, which are consistent with our intuition. Under the meaning of
statistic inversion by S. Takagi \cite{st86}, in odd dimensions, the observer
would feel a Bose-Einstein distribution for the Fermionic field and a
Fermi-Dirac distribution for the bosonic field. This counter-intuitive
phenomenon is interesting but it has not been fully understood because of the
relativity of the notion of acceleration. Some studies relevant to the
statistical inversion by the acceleration have been discussed, including
$2$-dimensional anyon field \cite{so17}, geometric phase \cite{hy12,fzz221},
and fisher information \cite{fz22}. Methods of optical lattice were used to
simulate the Unruh effect in different spacetime dimensions, and evident
statistical inversion from Fermi-Dirac to Bose-Einstein in $2$-dimensional
optical lattice with fermion gas might be observed, which could be a future
method to detect the statistic inversion experimentally \cite{rlc17,klc18}. A
recent study investigated the thermal nature of the Unruh effect in arbitrary
dimensions using open quantum systems \cite{bp02,kb20} coupled to massless
Minkowski vacuum \cite{abg21}, in which the relationship between the Unruh
effect and the thermal bath was also explored. In particular, it was pointed
out that the case for the massive field is quite different from the massless
field because the statistical factor would disappear \cite{st86,abg21}.

As far as we know, all studies about statistical inversion are restricted to a
single-atom system, and the entanglement behavior between two-atom systems in
any spacetime dimension has not been investigated up to now. In this paper,
based on Gorini-Kossakowski-Linblad-Sudarshan master equation \cite{bp02}, we
will study entanglement dynamics of two-atom system in different dimensions.
Meanwhile, the different initial states including the product and entangled
states for the two-atom system are considered, and the massless and massive
fields for the vacuum are also discussed in our paper.

This paper is organized as follows. In the second section, we investigate the
model of quantum field theory for the Unruh effect and the theory of open
quantum systems. We also give the concrete Wightman functions and their
Fourier transform in any spacetime dimensions and analyze their behavior. This
is followed in the third section by the evolution equations and concrete
behavior in different parameter conditions for the accelerated two-atom
system. The possible anti-Unruh effect is discussed in this section. In\ the
fourth section, we consider the case for the massive field and compare it with
the massless field case. Finally, we summarize and give all conclusions in the
fifth section. In all calculations of this paper, we take the natural units
$\hbar=c=k_{B}=1$, where $\hbar$ is the reduced Planck constant, $c$ is the
speed of light, and $k_{B}$ is the Boltzmann constant.

\section{Wightman functions and power spectrum}

In this section, we will discuss the Wightman functions and their Fourier
transform (power spectrum) for the massless field, and analyze their concrete
function form, and behavior with the change of $\omega$.

\subsection{Wightman functions}

The Unruh-DeWitt detector is an idealized model that captures all important
features of quantum field theory for describing the Unruh effect. Consider the
detector using a two-level atom with ground $|g\rangle$ and excited
$|e\rangle$ states which are separated by an energy gap $\omega$.

The Wightman function (sometimes also called correlation functions or
two-point functions) for the massless Minkowski field can be written as
\cite{st86}
\begin{equation}
G({x},{x^{^{\prime}}})={{\mathcal{C}}_{D}}{[(-1)({(t-t^{^{\prime}}-i\epsilon
)}^{2}-{|{x}-{x^{^{\prime}}}|}^{2})]}^{-(D-2)/2}, \label{mwf}%
\end{equation}
where $D$ represents the spacetime dimensions, $\epsilon\rightarrow0^{+}$,
${\mathcal{C}}_{D}=\frac{\Gamma{((D-2)/2)}}{4\pi^{(D/2)}}$ and $\Gamma$ stands
for the Gamma function. When the two atoms accelerate, their trajectories can
be expressed as%
\begin{align}
t_{1}(\tau)  &  =\frac{1}{a}\sinh{a\tau},x_{1}^{1}(\tau)=\frac{1}{a}%
\cosh{a\tau},x_{1}^{2}=x_{1}^{3}=\cdots=x_{1}^{D-2},x_{1}^{D-1}=0,\nonumber\\
t_{2}(\tau)  &  =\frac{1}{a}\sinh{a\tau},x_{2}^{1}(\tau)=\frac{1}{a}%
\cosh{a\tau},x_{2}^{2}=x_{2}^{3}=\cdots=x_{2}^{D-2},x_{2}^{D-1}=L, \label{tat}%
\end{align}
where $L$ is the fixed separation between two atoms along the ($D-1$)
coordinate. For this, the spacetime dimension should be $D\geq3$ in our consideration.

If the field state satisfies the KMS condition, one has \cite{st86,gmr16}
\begin{equation}
G^{(\alpha\varrho)}(\tau,\tau^{^{\prime}})=G^{(\alpha\varrho)}(\tau
-\tau^{^{\prime}}), \label{kms}%
\end{equation}
which implies that the Wightman function is stationary and depends only on the
difference between its two arguments. Inserting Eqs. (\ref{tat}) and
(\ref{kms}) into Eq. (\ref{mwf}), the diagonal components are obtained as
\begin{equation}
G^{\left(  11\right)  }{({x},{x^{^{\prime}}})}=G^{\left(  22\right)  }%
{({x},{x^{^{\prime}}})}={{\mathcal{C}}_{D}}\left(  \frac{{a}}{2i}\right)
^{D-2}\left[  {\sinh}\left(  \frac{{{a\tau}}}{2}{{-i\epsilon}}\right)
\right]  ^{-(D-2)} \label{wgc}%
\end{equation}
and the off-diagonal components are obtained as
\begin{equation}
G^{\left(  12\right)  }{({x},{x^{^{\prime}}})}=G^{\left(  21\right)  }%
{({x},{x^{^{\prime}}})}={\mathcal{C}}_{D}\left(  \frac{{a}}{2i}\right)
^{D-2}\left[  {{\sinh}}\left(  {{{\frac{{{a\tau}}}{2}-i\epsilon}}}\right)
{^{2}-}\frac{{{a^{2}}{L^{2}}}}{4}\right]  ^{-(D-2)/2} \label{wnc}%
\end{equation}

\subsection{Power spectrum}

The power spectrum is actually the Fourier transform of Wightman functions,
and it has four components in total. We discuss them with the diagonally
components and off-diagonally components, respectively.

\subsubsection{Diagonal components}

The power spectrum of (\ref{wgc}) can be expressed as
\begin{equation}
{\mathcal{G}}^{(a)}={{\mathcal{C}}_{D}}\left(  \frac{{a}}{2i}\right)
^{D-2}\int_{-\infty}^{\infty}d\tau e^{-i\omega\tau}\sinh\left(  {\frac
{{{a\tau}}}{2}-i\epsilon}\right)  ^{-(D-2)}, \label{pwd}%
\end{equation}
where ${\mathcal{G}}^{(a)}\equiv{\mathcal{G}}^{(11)}={\mathcal{G}}^{(22)}$. We
slightly downward the integration contour by $i\pi/a$ in such a way that the
integrand does not blow up in the singularity along the contour. Then, we get
\begin{equation}
{\mathcal{G}}^{(a)}={{\mathcal{C}}_{D}}\left(  \frac{{a}}{2i}\right)
^{D-2}\int_{-\infty-\frac{i\pi}{2a}}^{\infty-\frac{i\pi}{2a}}d\tau
e^{-i\omega\tau}\sinh\left(  {\frac{{{a\tau}}}{2}}\right)  ^{-(D-2)}.
\end{equation}
Make a variable substitution by $z=e^{a\tau+i\pi}$, and obtain
\begin{equation}
{\mathcal{G}}^{(a)}={\mathcal{C}}_{D}\frac{1}{a}\int_{0}^{\infty}%
dz\frac{z^{\frac{i\omega}{a}+\frac{D-2}{2}-1}}{{(1+z)}^{D-2}}.
\end{equation}
Using the Beta function \cite{jz07}
\begin{equation}
B(p,q)=\int_{0}^{\infty}dz\frac{z^{p-1}}{{(1+z)}^{p+q}}=\frac{\Gamma
{(p)}\Gamma{(q)}}{\Gamma{(p+q)}},
\end{equation}
and letting $p=\frac{i\omega}{a}+\frac{D-2}{2}$, $q=\frac{-i\omega}{a}%
+\frac{D-2}{2}$, we obtain
\begin{equation}
{\mathcal{G}}^{(a)}={\mathcal{C}}_{D}e^{-\frac{\pi\omega}{a}}\frac{1}{a}%
\frac{\Gamma{(\frac{i\omega}{a}+\frac{D-2}{2})}+\Gamma{(\frac{-i\omega}%
{a}+\frac{D-2}{2})}}{\Gamma{(D-2)}}.
\end{equation}
By the formula ${\Gamma{(x+iy)}}^{2}={\Gamma{(x+iy)}}\Gamma{(x-iy)}$
\cite{jz07}, the power spectrum is simplified into
\begin{equation}
{\mathcal{G}}^{(a)}={\mathcal{C}}_{D}\frac{1}{a}\frac{{|\Gamma{(\frac{D-2}%
{2}+\frac{i\omega}{a})}|}^{2}}{\Gamma{(D-2)}}. \label{aaa}%
\end{equation}
In order to present the power spectrum explicitly, the Euler's formulae
${|\Gamma{(ix)}|}^{2}=\pi/{(x\sinh{(\pi x)})}$, ${|\Gamma{(1/2+ix)}|}^{2}%
=\pi/\cosh{(\pi x)}$, and the property of Gamma function $\Gamma
(x+1)=(x+1)\Gamma(x)$ \cite{jz07} are used to obtain the following form,
\begin{equation}
{\mathcal{G}}^{(a)}=2\pi{\mathcal{C}}_{D}\frac{1}{\Gamma(D-2)}%
\begin{cases}
\frac{a^{D-2}}{\omega}\frac{1}{e^{\frac{2\pi\omega}{a}}-1}\prod\limits_{l=0}%
^{(D-4)/2}[l^{2}+{(}\frac{{\omega}}{a}{)}^{2}],D\quad is\quad even\\
a^{D-3}\frac{1}{e^{\frac{2\pi\omega}{a}}+1}\prod\limits_{l=0}^{(D-5)/2}%
[{(l+\frac{1}{2})}^{2}+{(\frac{{\omega}}{a})}^{2}],D\quad is\quad odd
\end{cases}
\label{pwf}%
\end{equation}

For the case of the spacetime dimension $D=3$, the continued product term
should be neglected. It is seen a Bose-Einstein factor in even dimensions and
a Fermi-Dirac factor in odd dimensions. This is consistent with the Takagi
statistical inversion: for Bosonic field, the accelerated observer sees a
Bose-Einstein distribution in even spacetime dimensions and a Fermi-Dirac
distribution in odd spacetime dimensions. Moreover, it can be inferred
directly from Eq. (\ref{pwf}) that the power spectrum ${\mathcal{G}}^{(a)}$
satisfies the KMS condition \cite{rk57,ms59}.
\begin{equation}
{\mathcal{G}}^{(a)}(\omega)=e^{\frac{-2\pi\omega}{a}}{\mathcal{G}}%
^{(a)}(-\omega), \label{psb}%
\end{equation}
which means that the ratio between the power spectra with positive and
negative frequency is a Boltzmann factor.

\begin{figure}[ptb]
\centering
\includegraphics[width=7in,height=4.5in]{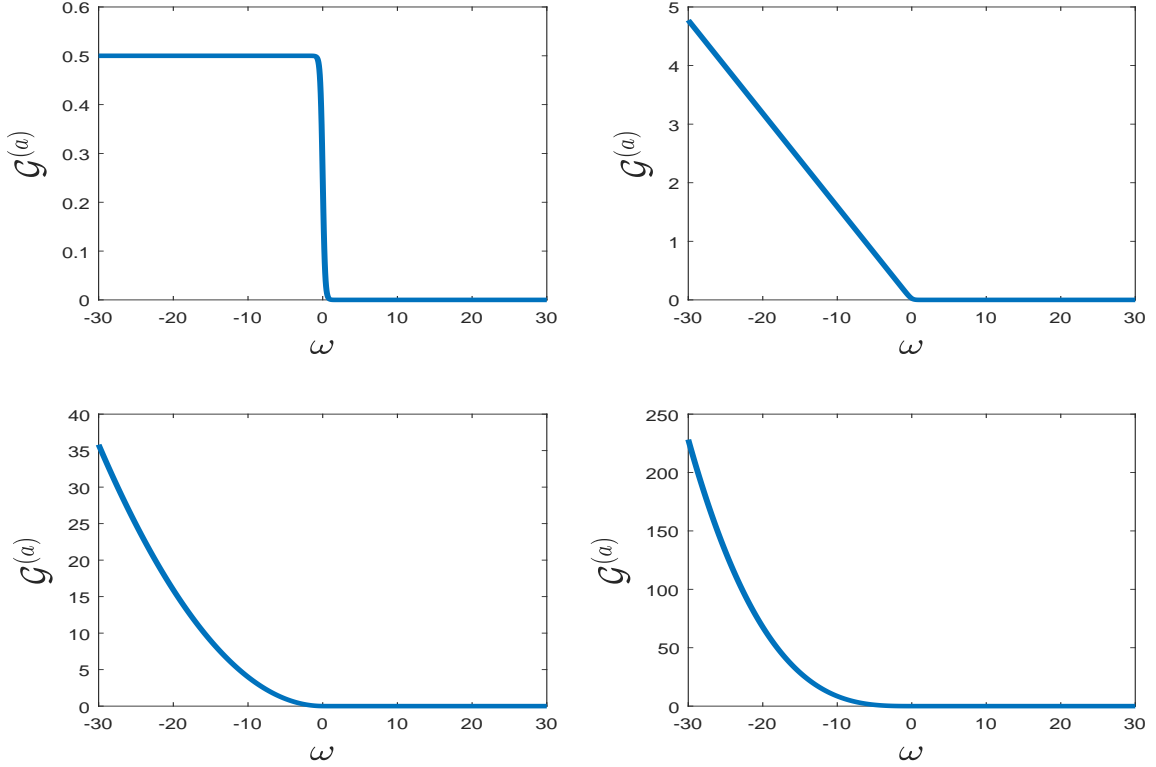}\caption{Behavior of
${\mathcal{G}}^{(a)}$ in different spacetime dimensions. The four subfigures
stand for $D = 3$, $D = 4$, $D = 5$, $D = 6$ respectively from left to right
and top to bottom.}%
\end{figure}

The behavior of the power spectra from $D=3$ to $D=6$ are depicted in Fig. 1
with the acceleration $a=1$. It is noticed that when $\omega>0$, the value for
${\mathcal{G}}^{(a)}$ is not zero though very small, and it equals to the
product of the corresponding negative frequency value and a Boltzmann factor
which restrains the corresponding numerical value, as in Eq. (\ref{psb}). From
Fig. 1 and Eq. (\ref{pwf}), it is seen that when $D=3$, ${\mathcal{G}}^{(a)}$
completely obeys the Fermi-Dirac distribution, which is the most special case;
when $D=4$, ${\mathcal{G}}^{(a)}$ is equivalent to a product between the
inverse of the factor $\omega$ and the Bose-Einstein factor; in particular,
when $D\geq5$, the behavior of ${\mathcal{G}}^{(a)}$ looks all similar since
the continued product term would dominate.

\subsubsection{Off-diagonal components}

Now we calculate the Fourier transform of the off-diagonal components in Eq.
(\ref{wnc}) with the expression as
\begin{equation}
{\mathcal{G}}^{(b)}={\mathcal{C}}_{D}\left(  {{\frac{a}{2i}}}\right)
^{D-2}\int_{-\infty}^{\infty}d\tau e^{-i\omega\tau}\left[  {\sinh}\left(
\frac{{{a\tau}}}{2}{{-i\epsilon}}\right)  {-}\frac{{{a^{2}}{L^{2}}}}%
{4}\right]  ^{(D-2)/2},
\end{equation}
where ${\mathcal{G}}^{(b)}\equiv{\mathcal{G}}^{(12)}={\mathcal{G}}^{(21)}$.
Similar to the calculation of (\ref{pwd}), slightly downward the integration
contour by $i\pi$, and we get
\begin{equation}
{\mathcal{G}}^{(b)}={\mathcal{C}}_{D}{({\frac{a}{i}})}^{D-2}\int_{-\infty
-i\pi}^{\infty-i\pi}d\tau e^{-i\omega\tau-\frac{D-2}{2}a\tau}{[(e^{-a\tau
}-e^{\gamma})(e^{-a\tau}-e^{\beta})]}^{-(D-2)/2},
\end{equation}
where $e^{\beta}=\frac{2+{a^{2}}{L^{2}}+aL\sqrt{{a^{2}}{L^{2}}+4}}{2}$ and
$e^{\gamma}=\frac{2+{a^{2}}{L^{2}}-aL\sqrt{{a^{2}}{L^{2}}+4}}{2}$. Let
$x\equiv a\tau+i\pi$ and we have
\begin{equation}
{\mathcal{G}}^{(b)}=a^{D-3}e^{\frac{-\pi\omega}{a}}{{\mathcal{C}}_{D}}%
\int_{-\infty}^{\infty}dx\frac{e^{-i{\frac{\omega}{a}}x-{\frac{D-2}{2}}x}%
}{{[{e^{-x}+e^{\beta}}]}^{\frac{D-2}{2}}{[{e^{-x}+e^{\gamma}}]}^{\frac{D-2}%
{2}}}.
\end{equation}
Make a further variable substitution, $t=\frac{e^{-\gamma}}{e^{x}+e^{-\gamma}%
}$, and the Fourier form becomes
\begin{equation}
{\mathcal{G}}^{(b)}=a^{D-3}{{\mathcal{C}}_{D}}e^{i\gamma\omega-{\frac{D-2}{2}%
}\beta}\int_{0}^{1}dt{(1-t)}^{\frac{D-2}{2}-i\omega-1}{[1-t(1-e^{\gamma-\beta
})]}^{-\frac{D-2}{2}}t^{\frac{D-2}{2}-1+i\omega}.
\end{equation}
Using one kind of integration expressions of hypergeometric function
\cite{jz07}
\begin{equation}
_{1}F^{2}(p,q;n,z)=\frac{\Gamma{(n)}}{\Gamma{(q)}\Gamma{(n-1)}}\int_{0}%
^{1}dt{(1-t)}^{n-q-1}{(1-tz)}^{-p}t^{q-1},
\end{equation}
and letting $p=\frac{D-2}{2}$, $q=\frac{D-2}{2}+i\omega$, $n=D-2$, and
$z=1-e^{\gamma-\beta}$, we obtain
\begin{equation}
{\mathcal{G}}^{(b)}=a^{D-3}{{\mathcal{C}}_{D}}e^{i\gamma\omega-{\frac{D-2}{2}%
}\beta}{\frac{{\Gamma{(\frac{D-2}{2}+i\omega)}{\Gamma{(\frac{D-2}{2}-i\omega
)}}}}{\Gamma{(D-2)}}}{_{1}F^{2}{(\frac{D-2}{2},\frac{D-2}{2}+i\omega
,D-2,1-e^{\gamma-\beta}).}}%
\end{equation}
Using the same method as we tackle (\ref{aaa}), an explicit expression for the
Fourier form of the off-diagonal components is given as
\begin{equation}
{\mathcal{G}}^{(b)}={\mathcal{T}}%
\begin{cases}
\frac{a^{D-2}}{\omega}\frac{1}{e^{\frac{2\pi\omega}{a}}-1}\prod\limits_{l=0}%
^{(D-4)/2}[l^{2}+{(\omega/a)}^{2}],D\quad is\quad even\\
a^{D-3}\frac{1}{e^{\frac{2\pi\omega}{a}}+1}\prod\limits_{l=0}^{(D-5)/2}%
[{(l+\frac{1}{2})}^{2}+{(\omega/a)}^{2}],D\quad is\quad odd
\end{cases}
\label{ncf}%
\end{equation}
where ${\mathcal{T}}=2^{D-5}\Gamma{(\frac{D-2}{2})}_{1}F^{2}{(\frac{D-2}%
{2},s;D-2,1-e^{\gamma-\beta})}{e^{{\gamma}p}}$, $s=\frac{i\omega}{a}%
+\frac{D-2}{2}$, $\beta=\log{[{\frac{1}{2}}{(2+{a^{2}}{L^{2}}+{aL}\sqrt
{{a^{2}}{L^{2}}+4})}]}$, $\gamma=-\beta$.

When the spacetime dimensions $D=3$, the continued product term should be
neglected. We can get directly from (\ref{ncf}) that it still obeys KMS
condition, ${\mathcal{G}}^{(b)}(\omega)=e^{\frac{-2\pi\omega}{a}}{\mathcal{G}%
}^{(b)}(-\omega)$. In particular, if we neglect all constant coefficients,
${\mathcal{G}}^{(b)}$ can be regraded as the multiplication of corresponding
${\mathcal{G}}^{(a)}$ and a hypergeometric function. \begin{figure}[ptb]
\centering
\includegraphics[width=7in,height=4.5in]{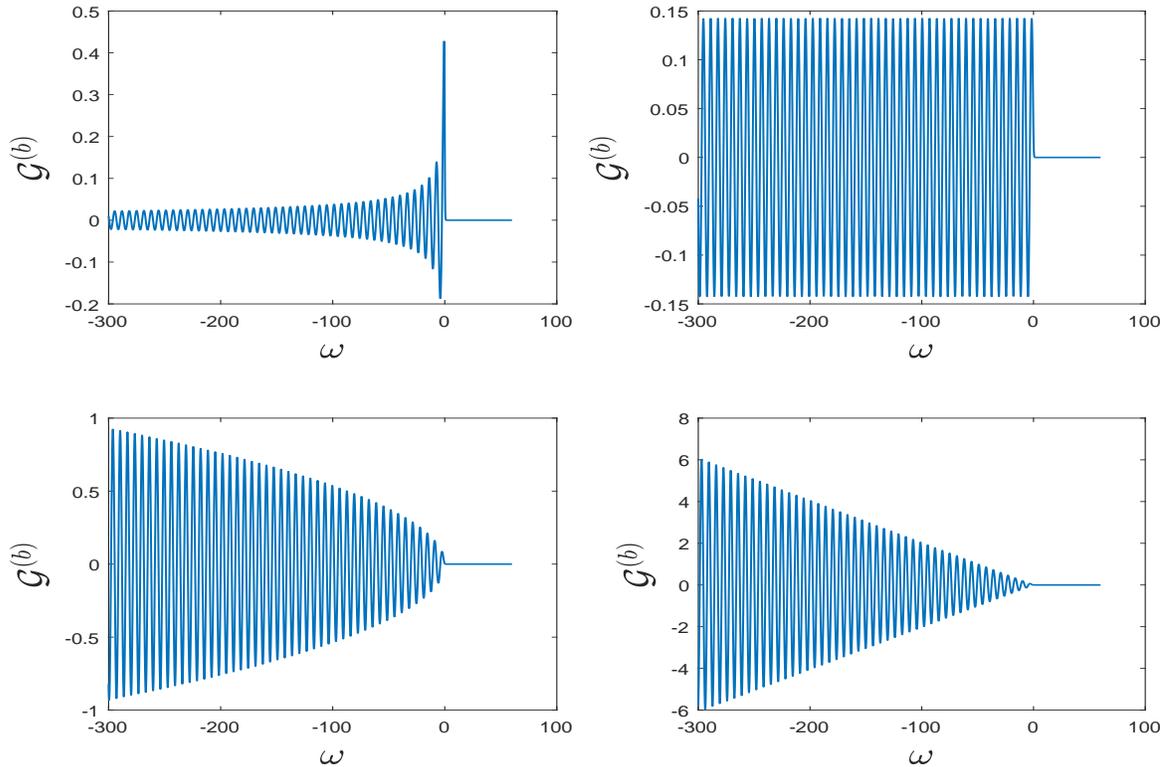}\caption{Behavior of
${\mathcal{G}}^{(b)}$ in different spacetime dimensions. The four subfigures
stand for $D=3$, $D=4$, $D=5$, $D=6$ respectively from left to right and top
to bottom.}%
\end{figure}

The behavior of the power spectra from $D=3$ to $D=6$ are depicted in Fig. 2
with the acceleration $a=1$ and the distance between two atoms $L=1$. It is
seen clearly from Fig. 2 that for $D=3$, the value for ${\mathcal{G}}^{(b)}$
will decrease by an oscillating form with $\omega$ decreasing from $0$ to
negative values; for $D=4$, ${\mathcal{G}}^{(b)}$ oscillates periodically but
the value is not decreasing for $\omega\leqslant0$; for $D=5$ and $D=6$,
${\mathcal{G}}^{(b)}$ will oscillate with an inverse trend comparing with the
case $D=3$ although the oscillation for $D=6$ is more violent than that for
$D=5$. The hypergeometric function term is responsible for the oscillation.
These oscillation can be understood form the expression of ${\mathcal{G}%
}^{(b)}$. For example, when $D=4$, the expression for ${\mathcal{G}}^{(b)}$
can be written as
\begin{equation}
\frac{1}{2\pi}\frac{1}{e^{\frac{2\pi\omega}{a}}-1}\frac{\sin{(\frac{2\omega
}{a}\sinh^{-1}{\frac{aL}{2}})}}{L\sqrt{1+a^{2}L^{2}/4}}.
\end{equation}
It is seen that when $\omega<0$, the Bose-Einstein factor is negligible and
the $\sin$ function would dominate the oscillation. Thus, the oscillation
amplitude would be not changed with the frequency $\omega$ for the constant
$a$ and $L$.

\section{Entanglement dynamics in any dimensions}

In the framework of open quantum systems \cite{bp02} and with the help of
Negativity, we study the relationship between entanglement dynamics and
spacetime dimensions in the case of the massless field for the vacuum. The
Unruh and anti-Unruh effects are also discussed in this section.

\subsection{Master equation}

Consider two atoms that consist of two energy levels for each one accelerating
in the Minkowski vacuum, and the total Hamiltonian for the interaction process
between accelerated atoms and the vacuum field can be written as
\begin{equation}
H=H_{A}+H_{F}+H_{I}.
\end{equation}
$H_{A}$ is the Hamiltonian of the two-atom system,
\begin{equation}
H_{A}=\frac{\omega}{2}\sigma_{3}^{(1)}+\frac{\omega}{2}\sigma_{3}^{(2)},
\end{equation}
where $\sigma_{i}^{(1,2)}=\sigma_{i}\otimes{\sigma_{0}}$, the superscripts
$(1,2)$ indicate the two different atoms, $\sigma_{i}(i=1,2,3)$ is the Pauli
matrices and $\sigma_{0}$ is the $2\times2$ unit matrix. $\omega$ is the
energy gap of each atom. $H_{F}$ is the Hamiltonian of the scalar fields that
represents the massless vacuum field. We assume the interaction Hamiltonian
$H_{I}$ is weak, with the following form,
\begin{equation}
H_{I}=\mu{[\sigma_{2}^{(1)}{\phi(t,{{x}}_{1})}+\sigma_{2}^{(2)}{\phi(t,{{x}%
}_{2})}],}%
\end{equation}
where $\mu$ is the coupling constant which is assumed to be small.

To make the calculation further, some approximations must be taken (see the
detailed discussion in Ref. \cite{abg21}). Under the Born-Markov approximation
\cite{bp02}, we can write the master equation describing the dissipative
dynamics of the two-atom system in the manner of
Gorini-Kossakowski-Lindblad-Sudarshan form,
\begin{equation}
\frac{\partial\rho{(\tau)}}{\partial\tau}=-i[H_{eff},\rho\left(  \tau\right)
]+\mathcal{D}{[\rho{(\tau)}]} \label{gkls}%
\end{equation}
where
\begin{equation}
H_{eff}=H_{A}-\frac{i}{2}\sum_{\alpha,\beta=1}^{2}\sum_{i,j=1}^{3}%
H_{ij}^{(\alpha\beta)}\sigma_{i}^{(\alpha)}\sigma_{j}^{(\beta)}%
\end{equation}
and
\begin{equation}
\mathcal{D}{[\rho{(\tau)}]}=\frac{1}{2}\sum_{\alpha,\beta=1}^{2}\sum
_{i,j=1}^{3}C_{ij}^{(\alpha\beta)}[2\sigma_{j}^{(\beta)}\rho\sigma
_{i}^{(\alpha)}-\sigma_{i}^{(\alpha)}\rho\sigma_{j}^{(\beta)}-\rho\sigma
_{i}^{(\alpha)}\sigma_{j}^{(\beta)}]
\end{equation}
From the master equation (\ref{gkls}), it is clear that the environment can
lead to dissipation and decoherence defined by the dissipator $\mathcal{D}%
{[\rho{(\tau)}]}$, and the coefficients $C_{ij}^{(\alpha\beta)}$ in the
dissipator is expressed as
\begin{equation}
C_{ij}^{(\alpha\beta)}=A^{(\alpha\beta)}\delta_{ij}-iB^{(\alpha\beta
)}\varepsilon_{ijk}\delta_{3k}-A^{(\alpha\beta)}\delta_{3i}\delta_{3j}%
\end{equation}
where
\begin{equation}%
\begin{split}
A^{(\alpha\beta)}  &  =\frac{\mu^{2}}{4}[\mathcal{G}^{(\alpha\beta)}%
{(-\omega)}+\mathcal{G}^{(\alpha\beta)}{(\omega)}]\\
B^{(\alpha\beta)}  &  =\frac{\mu^{2}}{4}[\mathcal{G}^{(\alpha\beta)}%
{(-\omega)}-\mathcal{G}^{(\alpha\beta)}{(\omega)}]
\end{split}
\label{abp}%
\end{equation}
and the concrete expression for $\mathcal{G}^{(\alpha\beta)}$ can be found in
Eqs. (\ref{pwf}) and (\ref{ncf}). $H_{ij}^{(\alpha\beta)}$ is the Hilbert
transform of corresponding power spectrum $G_{ij}{}^{(\alpha\beta)}$,
\begin{equation}
\mathcal{K}^{(\alpha\beta)}{(\lambda)}=\frac{1}{{\pi}i}P\int_{-\infty}%
^{\infty}d{\omega}\frac{G_{ij}^{(\alpha\beta)}{(\omega)}}{\omega-\lambda},
\end{equation}
where $P$ denoting the Cauchy principle value. Moreover, in this paper we just
consider the effect of environment (the vacuum field) on quantum entanglement
between two accelerated atoms in arbitrary dimensions, so the Hamiltonian for
any single atom and vacuum contribution terms can be neglected, and we just
need to consider the effect of dissipator $\mathcal{D}{(\rho{(\tau)})}$.

\subsection{Measure for two-atom quantum entanglement}

For convenience, we choose four Dicke states \cite{rhd54} as the bases for the
expression of density matrix,
\begin{align}
|e\rangle &  =|e_{1}\rangle\otimes|e_{2}\rangle,\nonumber\\
|s\rangle &  =(|g_{1}\rangle\otimes|e_{2}\rangle+|g_{2}\rangle\otimes
|e_{1}\rangle)/\sqrt{2},\label{dsb}\\
|g\rangle &  =|g_{1}\rangle\otimes|g_{2}\rangle,\nonumber\\
|a\rangle &  =(|g_{1}\rangle\otimes|e_{2}\rangle-|g_{2}\rangle\otimes
|e_{1}\rangle)/\sqrt{2}.\nonumber
\end{align}
The density matrix of the system can be written as
\begin{equation}
\rho{(t)}=\left[  {%
\begin{array}
[c]{cccc}%
\rho_{ee}{(t)} & \rho_{eg}{(t)} & 0 & 0\\
\rho_{eg}^{\ast}{(t)} & \rho_{gg}{(t)} & 0 & 0\\
0 & 0 & \rho_{ss}{(t)} & 0\\
0 & 0 & 0 & \rho_{aa}{(t)}%
\end{array}
}\right]  \label{dms}%
\end{equation}
where $\rho_{IJ}=\langle I|\rho|J\rangle$, $I,J=e,s,g,a$. We take Negativity
as measurement in order to define the entanglement amount of the two-atom system.

Negativity is an important measure for entanglement, it is defined
\cite{vw02,zsl98} according to
\begin{equation}
\mathcal{N}=\max{\left\{  0,-2\sum_{i}\mu_{i}\right\}  }%
\end{equation}
where $\mu_{i}$ are the eigenvalues of the partially transposition
\cite{ap96,hhh01} of the density matrix $\rho$ of two-body system. It is not
hard to confirm that $\mathcal{N}=0$ for untangled states of atoms and
$\mathcal{N}=1$ for maximally entangled states of atoms. Using the density
matrix (\ref{dms}), the Negativity is obtained as
\begin{equation}
\mathcal{N}=\max{\left\{  0,\mathcal{N}_{1},\mathcal{N}_{2}\right\}
}\label{ndm}%
\end{equation}
where
\begin{align}
\mathcal{N}_{1} &  =\sqrt{\mathcal{C}_{1}\mathcal{C}_{1}^{+}+{(\rho_{gg}%
+\rho_{ee})}^{2}}-{(\rho_{gg}+\rho_{ee}),}\nonumber\\
\mathcal{N}_{2} &  =\sqrt{\mathcal{C}_{2}\mathcal{C}_{2}^{+}+{(\rho_{aa}%
+\rho_{ss})}^{2}}-{(\rho_{aa}+\rho_{ss}),}\nonumber\\
\mathcal{C}_{1} &  =|\rho_{aa}-\rho_{ss}|-2\sqrt{\rho_{gg}-\rho_{ee}},\\
\mathcal{C}_{2} &  =2|\rho_{ge}{(t)}|-(\rho_{ss}({t})+\rho_{aa}{(t)}%
),\nonumber\\
\mathcal{C}_{1}^{+} &  =|\rho_{aa}-\rho_{ss}|+2\sqrt{\rho_{gg}-\rho_{ee}%
},\nonumber\\
\mathcal{C}_{2}^{+} &  =2|\rho_{ge}{(t)}|+(\rho_{ss}({t})+\rho_{aa}%
{(t)}).\nonumber
\end{align}

\subsection{Entanglement change for initial product states}

Combining (\ref{dms}) with (\ref{gkls}), we can get a set of differential
equations
\begin{align}
\rho_{gg}^{\prime} &  =-4(A_{a}-B_{a})\rho_{gg}+2(A_{a}+B_{a}-A_{b}-B_{b}%
)\rho_{aa}+2(A_{a}+B_{a}+A_{b}+B_{b})\rho_{ss},\nonumber\\
\rho_{ee}^{\prime} &  =-4(A_{a}+B_{a})\rho_{ee}+2(A_{a}-B_{a}-A_{b}+B_{b}%
)\rho_{aa}+2(A_{a}-B_{a}+A_{b}-B_{b})\rho_{ss},\nonumber\\
\rho_{aa}^{\prime} &  =-4(A_{a}-A_{b})\rho_{aa}+2(A_{a}-B_{a}-A_{b}+B_{b}%
)\rho_{gg}+2(A_{a}+B_{a}-A_{b}-B_{b})\rho_{ee},\label{des}\\
\rho_{ss}^{\prime} &  =-4(A_{a}+A_{b})\rho_{ss}+2(A_{a}-B_{a}+A_{b}-B_{b}%
)\rho_{gg}+2(A_{a}+B_{a}+A_{b}+B_{b})\rho_{ee},\nonumber\\
\rho_{ge}^{\prime} &  =-4A_{a}\rho_{ge},\qquad\rho_{eg}^{\prime}=-4A_{a}%
\rho_{eg},\nonumber
\end{align}
where $\rho_{IJ}^{\prime}=\frac{\partial\rho_{IJ}{(\tau)}}{\partial\tau}$, and
the parameters $A$ and $B$ are defined in Eq. (\ref{abp}) with the
corresponding concise signs here, $A_{a}\equiv A^{(11)}=A^{(22)}$,
$A_{b}\equiv A^{(12)}=A^{(21)}$, $B_{a}\equiv B^{(11)}=B^{(22)}$, and
$B_{b}\equiv B^{(12)}=B^{(21)}$.

In order to study the generation of entanglement, we choose the initial state
to be a product state $|10\rangle$, without loss of generality. From Eq.
(\ref{dsb}), it is easy to deduce that $\rho_{eg}=\rho_{ge}$ remains zero
during the whole process, which leads to the result that $\mathcal{N}_{2}$ in
Eq. (\ref{ndm}) is always negative. Thus, the Negativity can be calculated
according to
\begin{equation}
\mathcal{N}{(\rho{(\tau)})}=\max{(0,\mathcal{N}_{1}{(\tau)})}\label{nmd}%
\end{equation}
From (\ref{des}) and (\ref{nmd}), we can get the expression for the derivative
value of Negativity $\mathcal{N}_{1}{({\tau})}$ at $\tau=0$,
\begin{equation}
\mathcal{N}_{1}^{\prime}{(0)}=k(4|A_{b}^{2}|-4\sqrt{A_{a}^{2}-B_{a}^{2}%
})\label{ndv}%
\end{equation}
where $k$ is a coefficient related to the spacetime dimensions $D$, the
acceleration $a$, the difference $\omega$ of energy levels of atom and the
separation between atoms $L$, and $k$ is always positive. If $\mathcal{N}%
_{1}^{\prime}{(0)}>0$, it is deduced that entanglement is produced.

In the following part, we will investigate how large it is for the region of
different parameters ($a,\omega,L$) in which entanglement could be generated
in different spacetime dimensions.\newline\begin{figure}[ptb]
\centering
\includegraphics[width=7in,height=4.5in]{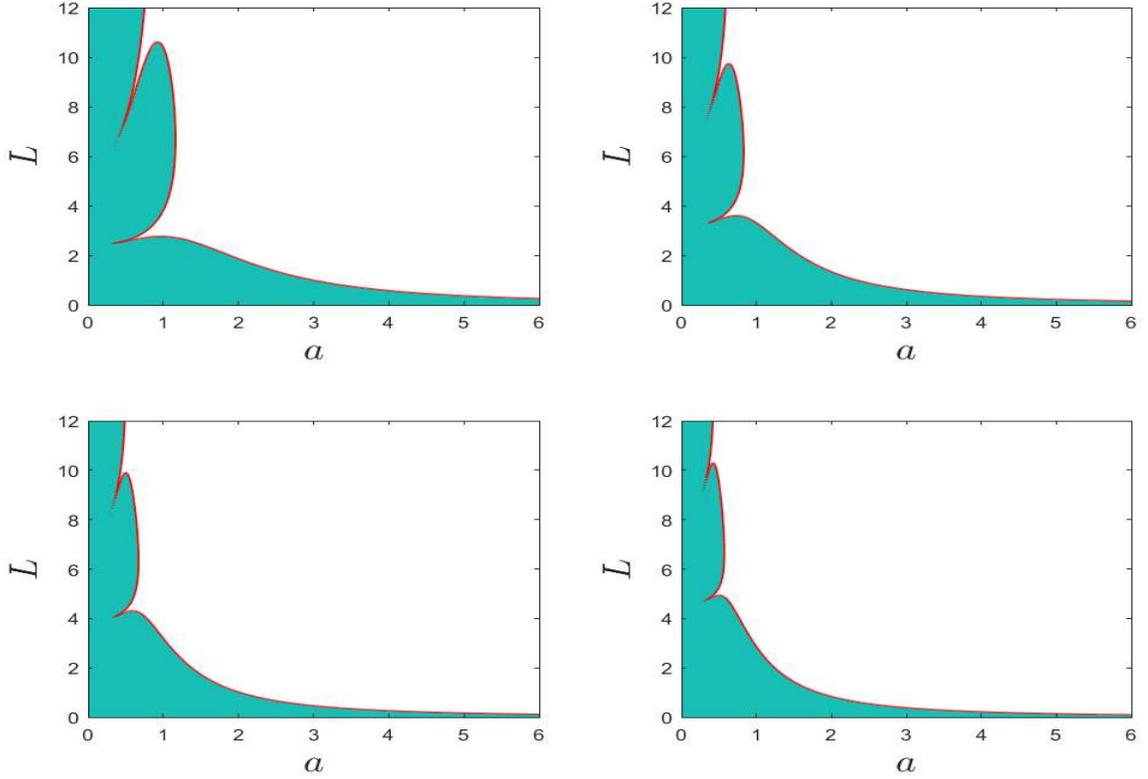}\caption{Entanglement
region in the $a-L$ plane. The four subfigures stand for $D=3$, $D=4$, $D=5$,
$D=6$ respectively from left to right and top to bottom.}%
\end{figure}

At first, we study the region where entanglement could be generated in the
$a-L$ plane with a fixed energy-level difference, $\omega=1$. Fig. 3 shows
this, and the area of presented entanglement region is estimated as $17.38$,
$13.63$, $11.88$, and $10.80$ for $D=3$, $D=4$, $D=5$, $D=6$, respectively. It
is stressed that the numerical values for the entanglement area don't have the
physical meaning, and it only indicates that the larger the area is, the
larger it is for the range of considered parameters with which entanglement
can be generated between atoms. It is not hard to find from Fig. 3 that the
area of the entanglement region in the $a-L$ plane would decrease with the
increasing spacetime dimensions. And we have calculated the area of
entanglement region for the same $D$ but different $\omega$, and found that
the larger $\omega$ is, the larger the area is.

\begin{figure}[ptb]
\centering
\includegraphics[width=7in,height=4.5in]{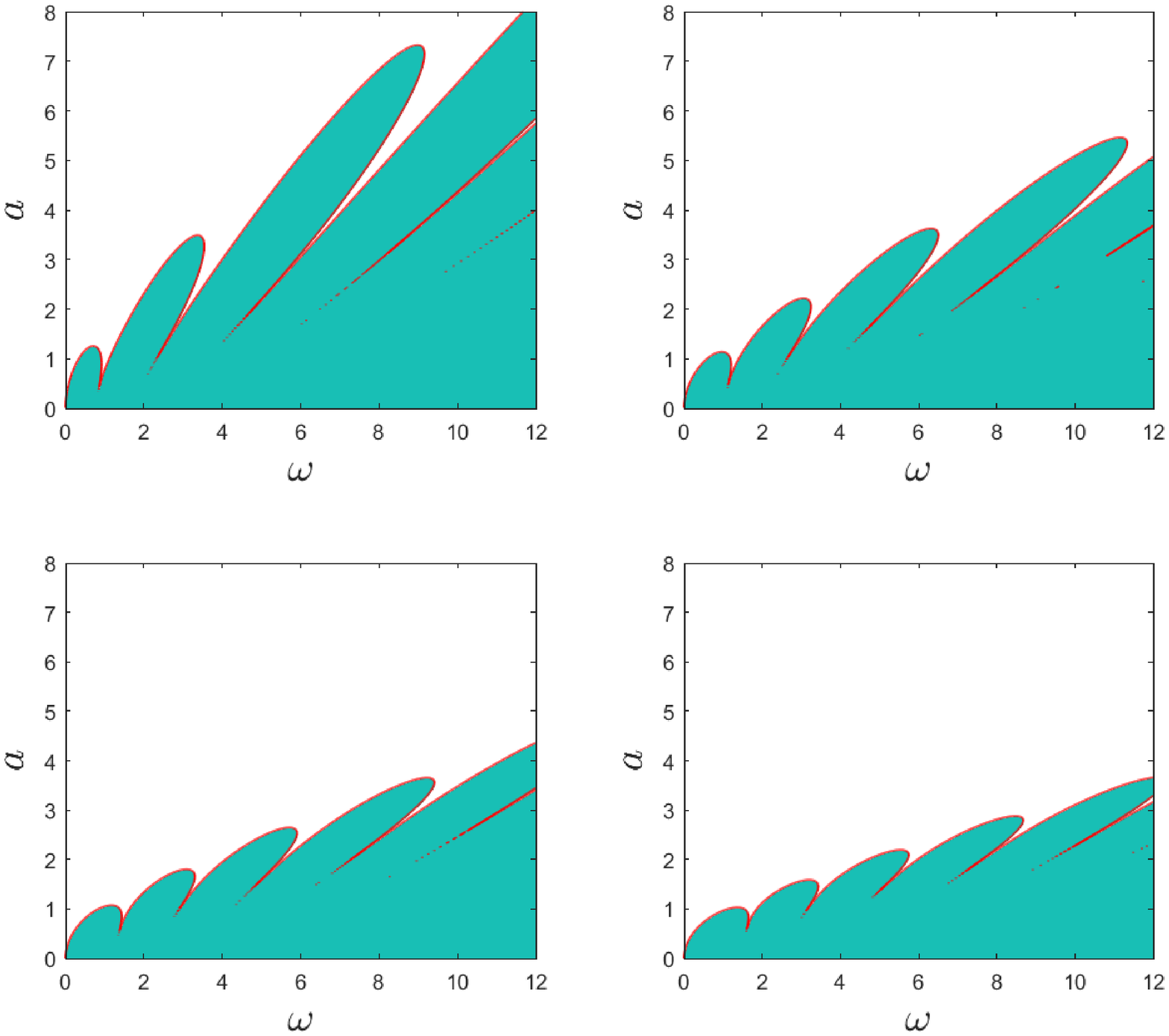}\caption{Entanglement
region in the $\omega-a$ plane. The four subfigures stand for $D=3$, $D=4$,
$D=5$, $D=6$ respectively from left to right and top to bottom.}%
\end{figure}

Then, we investigate the entanglement-generated region in the $\omega-a$ plane
with the separation between atoms fixed as $L=3$. As presented in Fig. 4, the
area of entanglement region is estimated as $55.39$, $37.96$, $30.15$, and
$25.86$ for $D=3$, $D=4$, $D=5$, $D=6$, respectively. Similar to the $a-L$
figure in Fig. 3, the area of the entanglement region in the $\omega-a$ plane
is decreasing with the increasing spacetime dimensions. And we have calculated
the area of entanglement region for the same $D$ but different $L$, and find
that the smaller $L$ is, the larger the area is.

\begin{figure}[ptb]
\centering
\includegraphics[width=7in,height=4.5in]{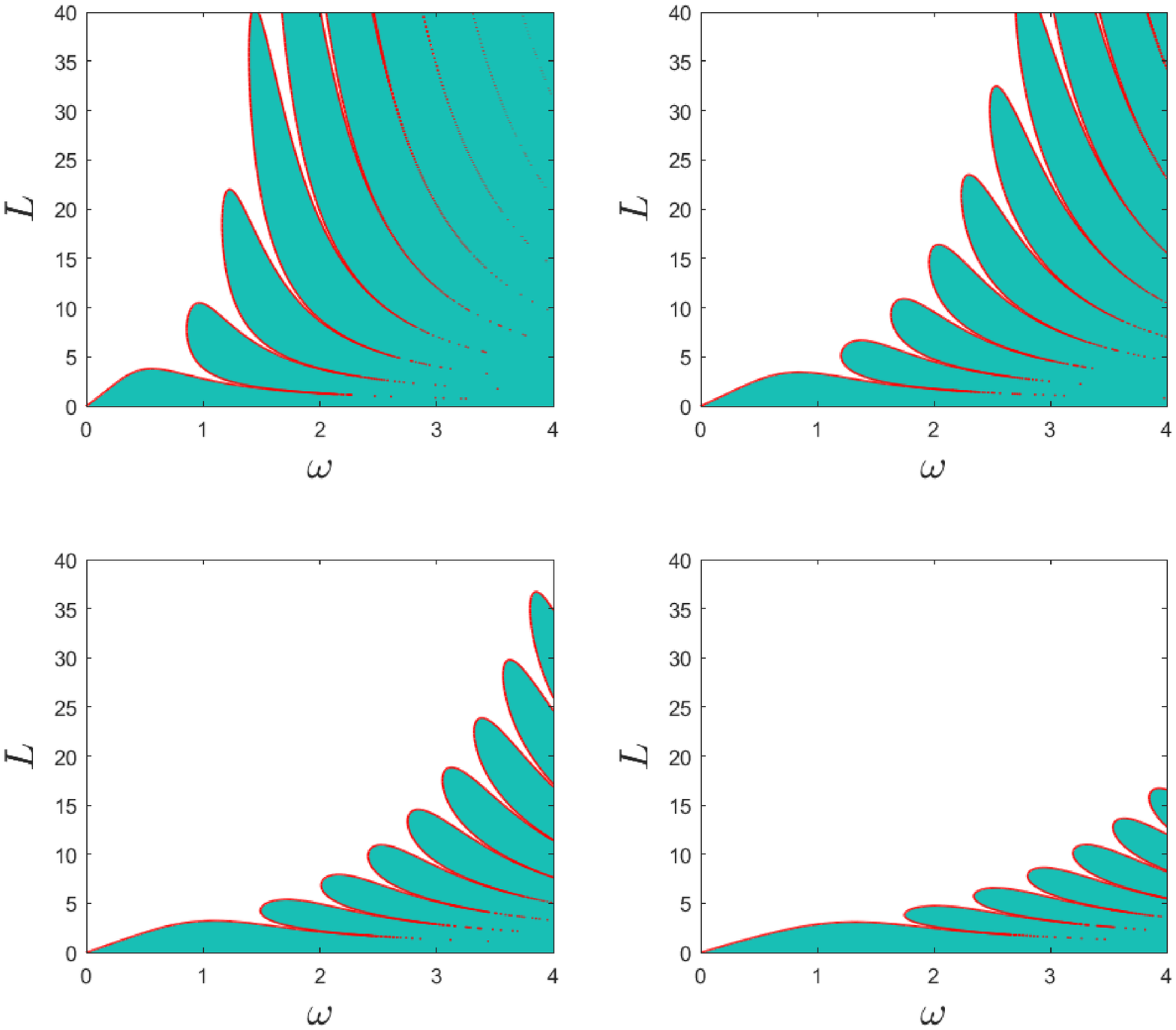}\caption{Entanglement
region in the $\omega-L$ plane. The four subfigures stand for $D=3$, $D=4$,
$D=5$, $D=6$ respectively from left to right and top to bottom.}%
\end{figure}

Finally, we fix $a=1$ and investigate the entanglement-generated region in the
$\omega-L$ plane. Similarly, we calculate the area of entanglement region as
$108.67$, $72.93$, $38.76$ and $22.06$ for $D=3$, $D=4$, $D=5$, $D=6$,
respectively, as presented in Fig. 5. It is also found that the area of
entanglement region in the $\omega-L$ plane is decreasing with the increasing
spacetime dimensions. And we have calculated the area of entanglement region
for the same $D$ but different $a$, and find that the smaller $a$ is, the
larger the area is.\newline

We make some discussions for Fig. 3, Fig. 4 and Fig. 5. From all the three
figures, it is seen that for larger spacetime dimensions, the range of the
parameters ($a,\omega,L$) in which entanglement can be generated is
decreasing. According to the calculation, we can give some interpretations.
From the theory of the master equation, the change of entanglement is derived
from a similar effect to that environment induced entanglement change. Here
the vacuum in different spacetime dimensions can be regarded as the
environment. When the spacetime dimensions increase, the environment would
become more sophisticated due to the addition of the spatial dimensions, which
can be manifested by our calculation to some extent. It is not hard to
understand that the entanglement region becomes smaller when the separation
between atoms or the acceleration increases, which means that the increased
parameters ($L,a$) would lead to the increase of difficulty for the generation
of entanglement. It is a little surprised that the entanglement region
increases when the energy level of the atom increases. More careful
investigation finds that the entangled value measured by Negativity would
decrease for increasing energy level of the atom although the area of the
entanglement region increases.

Furthermore, if we combine KMS condition (\ref{psb}) with Eq. (\ref{ndv}),
another condition for the entanglement region can be expressed as
\begin{equation}
\left\vert (1+e^{\frac{2\pi\omega}{a}}){\mathcal{G}}^{(b)}{(-\omega
)}\right\vert -2e^{\frac{\pi\omega}{a}}{\mathcal{G}}^{(a)}{(-\omega)}>0,
\label{ner}%
\end{equation}
which leads to a more refined result that the entanglement region is not
simply connected. For example, we fix $D=4,\omega=1$, and check the partially
amplified entanglement region in the $a-L$ plane, as given in Fig. 6. It is
found that there are many small holes where entanglement would not be
generated. This reason is that there is no evident boundary between
entanglement region and no-entanglement region, which can be deduced by the
fact that the ${\mathcal{G}}^{(b)}{(-\omega)}$ term in (\ref{ner}) is
oscillating with $a$ or $L$ (such oscillating behaviors cannot be presented
using the figures but it can be observed only by the analyzed expressions in
the second section) but ${\mathcal{G}}^{(a)}{(-\omega)}$ term is monotonous
with $a$ or $L$. \begin{figure}[ptb]
\centering
\includegraphics[width=5in,height=3in]{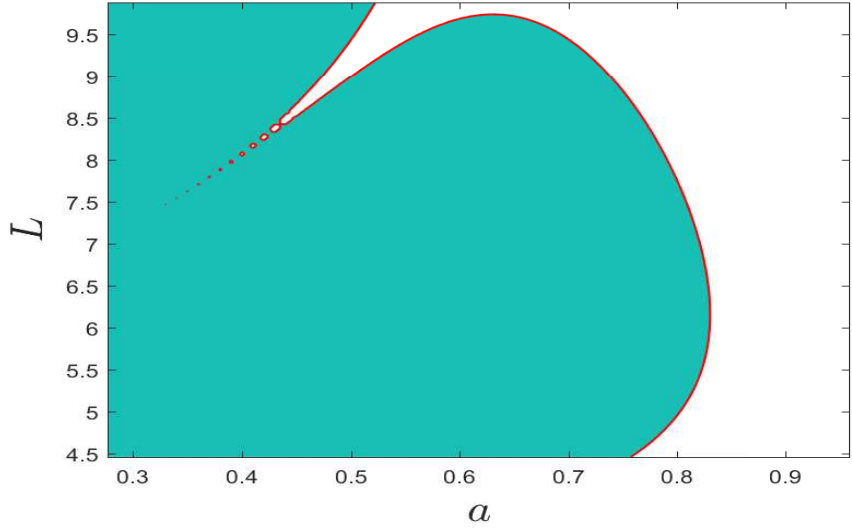}\caption{Partially
amplified figure of the $D=4$ case of Fig. 3.}%
\end{figure}

Now we discuss the change of entanglement generated after atoms are
accelerated with time in different spacetime dimensions. Fig. 7 presents this
evolution by calculating the amount of Negativity with the parameters $L=1$,
$\omega=1$, and $a=1$. It is seen clearly that when the spacetime dimension is
smaller, entanglement is generated more quickly but decays also more quickly.
Meanwhile, when the spacetime dimension decreases, the maximal value of
entanglement increases, but the entanglement duration time (this is estimated
by the half-width of the evolution curves) decreases. \begin{figure}[ptb]
\centering
\includegraphics[width=5in,height=3in]{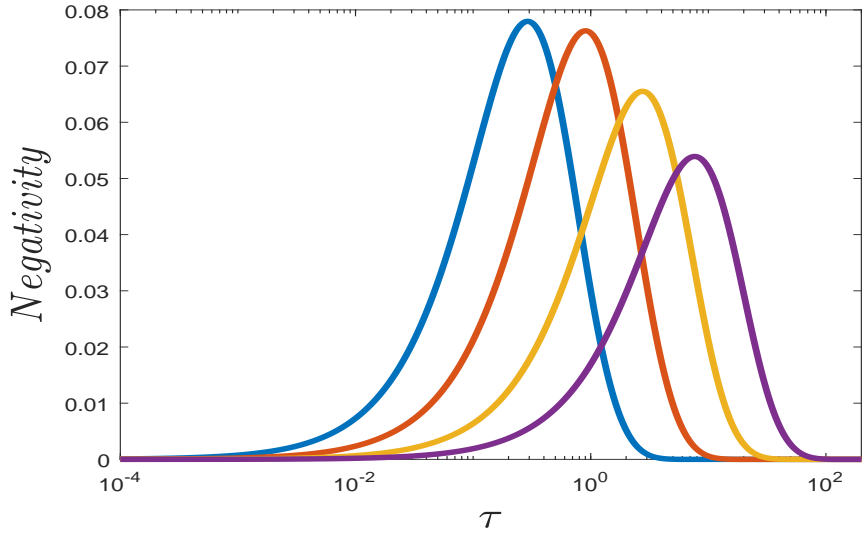}\caption{Entanglement
evolutionary process over time. Blue, red, yellow and purple lines stand for
$D=3$, $D=4$, $D=5$, $D=6$, respectively.}%
\end{figure}

Fig. 8 presents the change of maximal values of generated entanglement using
Negativity with $a$ throughout the whole evolved process. We plot different
results for different $L$. It is seen that for small $L$, the maximum of
Negativity will decrease monotonously with increasing acceleration, and for
larger spacetime dimensions, entanglement would decrease to zero more quickly.
When $L$ increases, the maximum of Negativity would not decrease monotonously
but increase first and decrease then to zero, and this phenomenon will happen
for smaller $L$ under smaller spacetime dimensions. The increase of
entanglement with the increasing acceleration $a$ looks like the phenomena
appeared in the case of anti-Unruh effect \cite{bmm16,lzy18}. We will not
discuss this here more detailed, and postpone it to the next section for the
case of the massive field. When $L$ takes larger values, the maximum of
Negativity will oscillate with increasing acceleration. This oscillation is
interesting but there is not an explicit interpretation for it, which might
deserve to be studied further in the future. Moreover, it is also noted that
the possible generated maximum of entanglement will decreases with the
increasing separation between atoms, as expected. \begin{figure}[ptb]
\centering
\includegraphics[width=7in,height=4.5in]{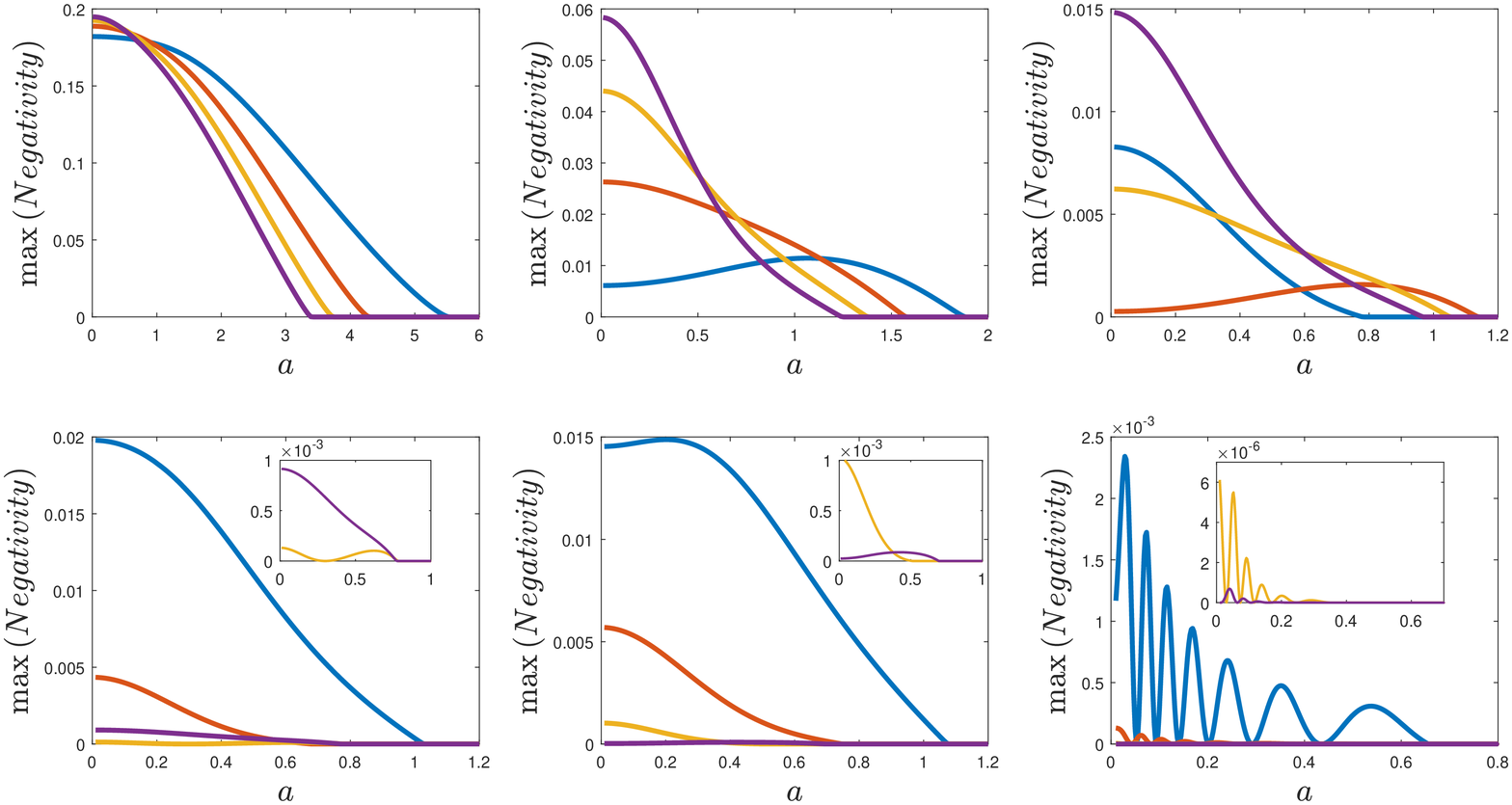}\caption{Changes of
maximal Negativity with acceleration for different atom separations. The six
subfigures stand for $L=0.3$, $L=2$, $L=3$, $L=4$, $L=4.4$, and $L=30$
respectively from left to right and top to bottom. Blue, red, yellow and
purple lines stand for $D=3$, $D=4$, $D=5$, and $D=6$, respectively.}%
\end{figure}

We also calculate the change of maximal values of generated entanglement using
Negativity with other parameters (e.g. $L$ or $\omega$) throughout the whole
evolved process. The results are similar to that presented in Fig. 8 and there
is not any novel behavior, so we would not discuss these here.

\subsection{Entanglement change for initial entangled states}

In this section, we study the change of entanglement with the initial
entangled states, $\alpha|10\rangle+\beta|01\rangle$ $\left(  \alpha,\beta
\neq0,\alpha^{2}+\beta^{2}=1\right)  $. Fig. 9 presents the change of
entanglement with time with the parameters $\omega=1$, $L=0.3$, $a=1$. As
shown in the left one of Fig. 9, the amount of entanglement will decrease
monotonously with time for initially maximally entangled states. In
particular, the larger the spacetime dimension $D$ is, the later it is for
time that entanglement disappears. When the initial entanglement is not big
enough, it will decrease first to zero, followed by a slight increase, and
then decrease to zero finally, as presented in the middle one of Fig. 9. It is
interesting for the small initial entanglement in the right one of Fig. 9, it
decreases to zero at first and then increases to a value larger than the
initial entanglement, and the generated maximal entanglement by accelerating
two atoms with an initial product state as in Fig. 7. It seems that the
initial small entanglement can boost the generated amount of entanglement by
the acceleration. This looks like an amplification mechanism for quantum
entanglement, but the amplification will be not valid if the initial amount of
entanglement is large enough, as in the middle one of Fig. 9.
\begin{figure}[ptb]
\centering
\includegraphics[width=7in,height=2.0in]{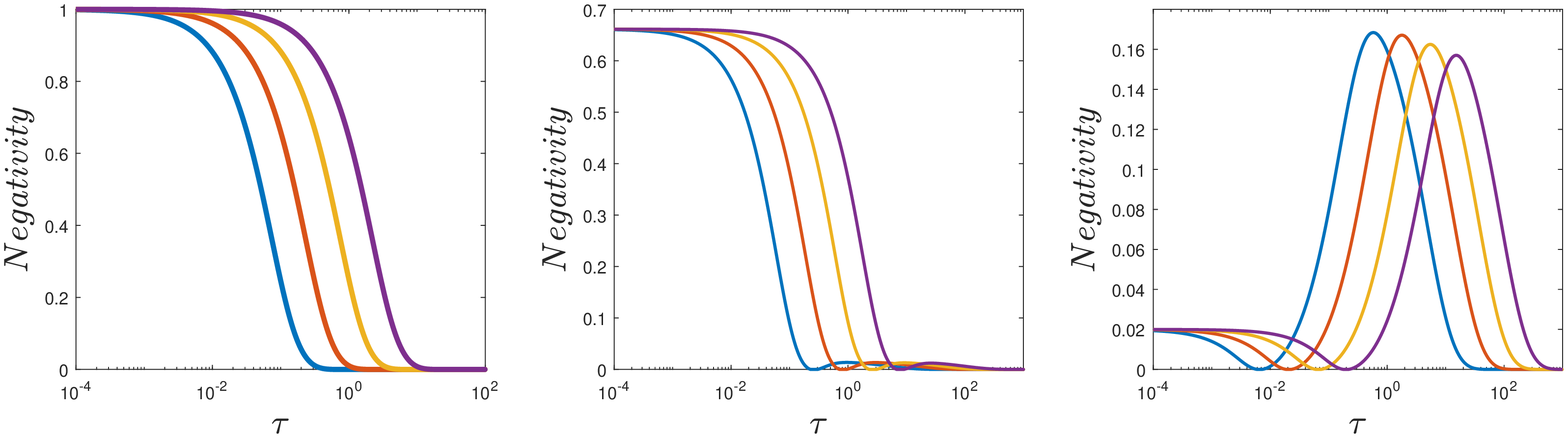}\caption{Entanglement
evolutionary process over time for different initial entangled states. The
three subfigures stand for $\alpha=1/\sqrt{2}$, $\alpha=1/{(2\sqrt{2})}$,
$\alpha=0.01$, respectively, from left to right.}%
\end{figure}

In the following, we discuss the change of entanglement for initial entangled
states in two different asymptotic cases, $L\rightarrow\infty$ and
$L\rightarrow0$. At first, we discuss the case with $L\rightarrow\infty$. In
this condition, ${\mathcal{G}}^{(b)}=0$, $A_{b}=B_{b}=0$. We can analytically
solve (\ref{des}) and obtain the corresponding results as
\begin{align}
\rho_{aa} &  =\frac{A_{1}^{2}-B_{1}^{2}+{A_{1}^{2}}e^{-8A_{1}\tau}-{B_{1}^{2}%
}e^{-8A_{1}\tau}-4A_{1}^{2}{\alpha\beta}{e^{-4A_{1}\tau}}+{2B_{1}^{2}%
}e^{-4A_{1}\tau}}{4A_{1}^{2}}\nonumber\\
\rho_{ee} &  =\frac{(A_{1}-B_{1}){e^{-8A_{1}\tau}}(-1+e^{4A_{1}\tau}%
)(A_{1}+B_{1}+A_{1}e^{4A_{1}\tau}-B_{1}e^{4A_{1}\tau})}{4A_{1}^{2}}\nonumber\\
\rho_{gg} &  =\frac{(A_{1}+B_{1}){e^{-8A_{1}\tau}}(-1+e^{4A_{1}\tau}%
)(A_{1}-B_{1}+A_{1}e^{4A_{1}\tau}+B_{1}e^{4A_{1}\tau})}{4A_{1}^{2}}\\
\rho_{ss} &  =\frac{A_{1}^{2}-B_{1}^{2}+{A_{1}^{2}}e^{-8A_{1}\tau}-{B_{1}^{2}%
}e^{-8A_{1}\tau}+4A_{1}^{2}{\alpha\beta}{e^{-4A_{1}\tau}}+{2B_{1}^{2}%
}e^{-4A_{1}\tau}}{4A_{1}^{2}}\nonumber
\end{align}
Substituting this into (\ref{ndm}), Negativity can be given in any time of the
evolution progress. It is found that Negativity will decrease rapidly to zero,
no matter what initial state or acceleration we choose.

Next, let $L\rightarrow0$, under which $A_{a}=A_{b}$, $B_{a}=B_{b}$. We can
solve (\ref{des}) analytically and obtain the corresponding results as
%\newpage%
\begin{equation}
\rho_{aa}=\frac{1}{2}{(\alpha-\beta)}^{2},
\end{equation}%
\begin{align}
\rho_{ee} &  =\frac{(1+2\alpha\beta){(A_{1}-B_{1})}^{2}}{6A_{1}^{2}+2B_{1}%
^{2}}\nonumber\\
&  -((1+2\alpha\beta)(-A_{1}+B_{1}+\sqrt{(A_{1}-B_{1})(A_{1}+B_{1})}%
))(A_{1}^{3}-A_{1}B_{1}^{2}+2B_{1}^{2}\sqrt{(A_{1}-B_{1})(A_{1}+B_{1}%
)})\nonumber\\
&  +((A_{1}-B_{1}){(A_{1}+B_{1})}^{3/2})e^{-4(2A_{1}+\sqrt{(A_{1}-B_{1}%
)(A_{1}+B_{1})})\tau}/(4B_{1}\sqrt{(A_{1}-B_{1})(A_{1}+B_{1})}(3A_{1}%
^{2}+B_{1}^{2}))\nonumber\\
&  +((1+2\alpha\beta)(A_{1}-B_{1}+\sqrt{(A_{1}-B_{1})(A_{1}+B_{1})}%
))(-A_{1}^{3}+A_{1}B_{1}^{2}+2B_{1}^{2}\sqrt{(A_{1}-B_{1})(A_{1}+B_{1}%
)}\nonumber\\
&  +((A_{1}-B_{1})(A_{1}+B_{1})^{3/2}))e^{-4(2A_{1}-\sqrt{A_{1}^{2}-B_{1}^{2}%
})\tau}/(4B_{1}\sqrt{(A_{1}-B_{1})(A_{1}+B_{1})}{(3A_{1}^{2}+B_{1}^{2})}),
\end{align}%
\begin{align}
\rho_{gg} &  =\frac{(1+2\alpha\beta){(A_{1}+B_{1})}^{2}}{6A_{1}^{2}+2B_{1}%
^{2}}\nonumber\\
&  -((1+2\alpha\beta)(A_{1}+B_{1}-\sqrt{(A_{1}-B_{1})(A_{1}+B_{1})}%
))(A_{1}^{3}-A_{1}B_{1}^{2}+2B_{1}^{2}\sqrt{(A_{1}-B_{1})(A_{1}+B_{1}%
)})\nonumber\\
&  +((A_{1}-B_{1}){(A_{1}+B_{1})}^{3/2})e^{-4(2A_{1}+\sqrt{(A_{1}-B_{1}%
)(A_{1}+B_{1})})\tau}/(4B_{1}\sqrt{(A_{1}-B_{1})(A_{1}+B_{1})}(3A_{1}%
^{2}+B_{1}^{2}))\nonumber\\
&  -((1+2\alpha\beta)(A_{1}+B_{1}+\sqrt{(A_{1}-B_{1})(A_{1}+B_{1})}%
))(-A_{1}^{3}+A_{1}B_{1}^{2}+2B_{1}^{2}\sqrt{(A_{1}-B_{1})(A_{1}+B_{1}%
)}\nonumber\\
&  +((A_{1}-B_{1})(A_{1}+B_{1})^{3/2}))e^{-4(2A_{1}-\sqrt{A_{1}^{2}-B_{1}^{2}%
})\tau}/(4B_{1}\sqrt{(A_{1}-B_{1})(A_{1}+B_{1})}{(3A_{1}^{2}+B_{1}^{2})}),
\end{align}%
\begin{align}
\rho_{ss} &  =\frac{(1+2\alpha\beta)(A_{1}^{2}-B_{1}^{2})}{2(3A_{1}^{2}%
+B_{1}^{2})}+\frac{(1+2\alpha\beta)(A_{1}^{2}+B_{1}^{2}-A_{1}\sqrt
{(A_{1}-B_{1})(A_{1}+B_{1})})e^{-4(2A_{1}-\sqrt{A_{1}^{2}-B_{1}^{2}})\tau}%
}{2(3A_{1}^{2}+B_{1}^{2})}\nonumber\\
&  +\frac{(1+2\alpha\beta)(A_{1}^{2}+B_{1}^{2}+A_{1}\sqrt{(A_{1}-B_{1}%
)(A_{1}+B_{1})})e^{-4(2A_{1}+\sqrt{(A_{1}-B_{1})(A_{1}+B_{1})})\tau}}%
{2(3A_{1}^{2}+B_{1}^{2})}.
\end{align}

\begin{figure}[ptb]
\centering
\includegraphics[width=5in,height=3in]{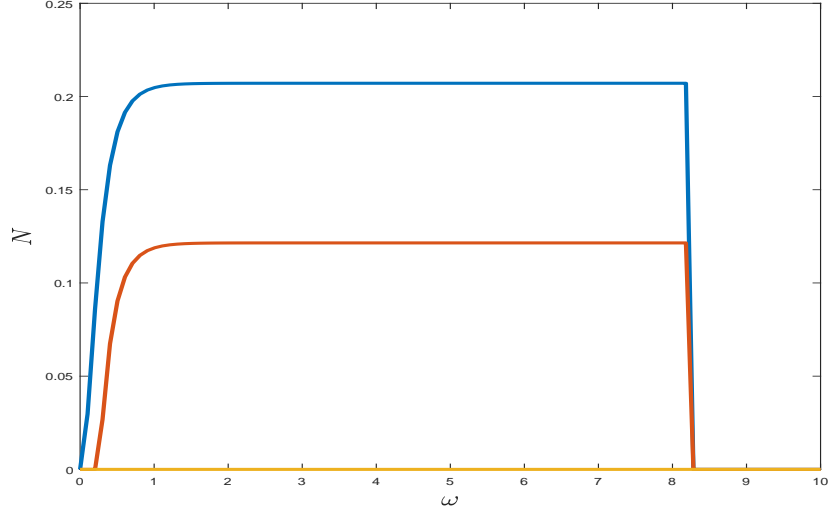}\caption{The change of N
with $\omega$ for different initial states, with $\alpha=0$ for the blue line,
$\alpha=0.1$ for the red line, and $\alpha=1/\sqrt{2}$ for the yellow line. }%
\end{figure}

Assume that when $\tau\rightarrow\infty$, the value for Negativity is $N$. The
change of $N$ with respect to $\omega$ is shown in Fig. 10 with the parameters
$a=1.4,D=3$. From this figure, it is found that for a general entangled state
initially (see the red line in Fig. 10), when $\omega$ is very small, the
Negativity will decrease to zero; after $\omega$ increases to some value, $N$
will grows rapidly to a fixed value, but this fixed value is related to the
initial states; when $\omega$ becomes larger, $N$ will decrease suddenly to
zero and keep zero for larger $\omega$. That is to say, if there is no
separation between two atoms, entanglement will not disappear within a certain
range of $\omega$, except for the initially maximally entangled state
($\alpha=\beta=\frac{1}{\sqrt{2}}$) where entanglement always remains at zero,
as presented with the yellow line in Fig. 10. In particular, we find that the
amount of generated entanglement by the acceleration is maximal if the initial
state is a product state ($\alpha=1$ or $\beta=1$), as presented with the blue
line in Fig. 10.

\section{Massive field}

In the last section, we investigate the dynamics of entanglement for two
accelerated atoms under the condition that the vacuum field is considered as
the massless field. In this section, we will turn to the consideration of a
massive field for the vacuum and discuss the influence of acceleration on
entanglement under different spacetime dimensions.

\subsection{Wightman function and power spectrum}

In $D$-dimensional Minkowski spacetime, the scalar field operator could be
expanded as
\begin{equation}
\phi(t,{x})=\int d^{D-1}{k}\frac{1}{\sqrt{2{\omega_{k}}{2\pi}^{D-1}}}[{a_{k}%
}e^{i{kx}-i{\omega_{k}}t}+{a_{k}^{\dagger}}e^{-i{kx}+i{\omega_{k}}%
t}],\label{sfo}%
\end{equation}
where ${a_{k}}$ (${a_{k}^{\dagger}}$) is the annihilation (creation) operator.
The expression for Wightman function is
\begin{equation}
G^{(\alpha\varrho)}(\Delta t)=\langle\phi(t_{\alpha}(t),{x}_{\alpha}%
(t))\phi(t_{\varrho}(t^{^{\prime}}),{x}_{\varrho}(t^{^{\prime}}))\rangle
,\label{mwfe}%
\end{equation}
where $\Delta t=t-t^{^{\prime}}$ is the difference of time. Inserting
(\ref{sfo}) into (\ref{mwfe}), one obtains
\begin{equation}
G^{(\alpha\varrho)}{(\Delta t)}=\frac{1}{(2\pi)^{D-1}}\int d^{D-1}k\frac
{1}{2\omega_{k}}e^{-i\omega_{k}{\Delta}t+i{kx}}.
\end{equation}
Here the integral of $d^{D-1}k$ can be regarded as the integral of the length
and the direction of the vector whose direction can be described by ($D-2$)
angular coordinates. After integrating all the angular variables, the Wightman
function becomes
\begin{equation}
G^{(\alpha\varrho)}{(\Delta t)}=\frac{1}{(4\pi)^{\frac{D-1}{2}}\Gamma
{(\frac{D-1}{2})}}\int_{0}^{\infty}{dk}\frac{k^{D-3}}{\omega_{k}}\frac
{\sin{(k{|\Delta{{x}}_{\alpha\varrho}|})}}{|\Delta{{x}}_{\alpha\varrho}%
|}e^{-i{\omega_{k}}{\Delta}t_{\alpha\varrho}}.
\end{equation}
Using the dispersion relationship $\omega_{k}^{2}=k^{2}+m^{2}$ to replace the
integration variable $k$ by ${\omega_{k}}$, we obtain
\begin{equation}
G^{(\alpha\varrho)}{(\Delta\tau)}=\int_{m}^{\infty}d{\omega_{k}}\frac
{(\sqrt{\omega_{k}^{2}-m^{2}})^{D-4}}{|{\Delta}{{x}_{\alpha\varrho}}|}%
\sin{(\sqrt{\omega_{k}^{2}-m^{2}}|{\Delta}{{x}_{\alpha\varrho}}|)}%
e^{-i{\omega_{k}}{\Delta}t_{\alpha\varrho}},\label{mwff}%
\end{equation}
where $|{\Delta}{{x}_{\alpha\varrho}}|=\sqrt{|{{x}_{\alpha}-{{x}_{\varrho
}^{\prime}}}|}$, ${\Delta}t_{\alpha\varrho}=t_{\alpha}-t_{\varrho}^{\prime}$.

Now inserting the trajectories (\ref{tat}) of two atoms into the expression
(\ref{mwff}), we obtain
\begin{equation}
G^{(11)}{(\Delta\tau)}=G^{(22)}{(\Delta\tau)}=\frac{m^{D-2}}{{(4\pi)}%
^{\frac{D-1}{2}}\Gamma{(\frac{D-1}{2})}}\int_{1}^{\infty}dx{(x^{2}-1)}%
^{\frac{D-3}{2}}e^{-i\frac{2}{a}mx\sinh{\frac{a\Delta\tau}{2}}}, \label{mde}%
\end{equation}
and
\begin{equation}
G^{(12)}{(\Delta\tau)}=G^{(21)}{(\Delta\tau)}=\frac{m^{D-3}}{L{(4\pi
)^{\frac{D-1}{2}}{\Gamma{(\frac{D-1}{2})}}}}\int_{1}^{\infty}dx{(x^{2}%
-1)}^{\frac{D-4}{2}}sin{(mL{(x^{2}-1)}^{\frac{1}{2}})}e^{-i{\frac{2}{a}%
}mx{\sinh{\frac{a\Delta\tau}{2}}}}. \label{moe}%
\end{equation}
Further, we can write the expressions (\ref{mde}) in the manner of $K$-Bessel
function \cite{jz07} as
\begin{equation}
G^{(11)}{(\Delta\tau)}=G^{(22)}{(\Delta\tau)}=\frac{m^{D-2}}{{(4\pi)}%
^{\frac{D-1}{2}}(i\frac{m}{a}{\sinh{(\frac{a\Delta\tau}{2})}}^{\frac{D-2}{2}%
}\Gamma{(\frac{1}{2})})}K_{\frac{D-2}{2}}\left(  {i\frac{2m}{a}\sinh
{(\frac{a\Delta\tau}{2})}}\right)  \label{mdem}%
\end{equation}
where $K$ represents modified Bessel functions of the second kind. But for the
off-diagonal terms of the Wightman function, only when $D=4$, they can be
written in the manner of $K$-Bessel function \cite{jz07},
\begin{equation}
G^{(12)}{(\Delta\tau)}=G^{(21)}{(\Delta\tau)}=\frac{m}{{(4\pi)}^{\frac{3}{2}%
}\Gamma{(\frac{3}{2})}}\frac{1}{\sqrt{L^{2}-\frac{4}{a^{2}}\sinh^{2}%
{(\frac{a\Delta\tau}{2})}}}K_{1}\left(  {m\sqrt{L^{2}-\frac{4}{a^{2}}\sinh
^{2}{(\frac{a\Delta\tau}{2})}}}\right)  . \label{moem}%
\end{equation}

It is found that the Wightman functions for the case considering the massive
field can be regarded as the multiplication of a $K$-Bessel function and the
Wightman functions for the case considering the massless field up to a
constant related to mass $m$. Thus, it is not easy to obtain the Fourier
transform of the Wightman functions for the case considering the massive field
with numerical methods because the absolute values of the argument of
$K$-Bessel functions are highly oscillatory. In order to acquire the power
spectrum in the case of the massive field, we rewrite the Fourier transform of
(\ref{mde}) as
\begin{equation}
{\mathcal{G}}^{(a)}=\frac{am^{D-3}}{{(4\pi)}^{\frac{D-1}{2}}\Gamma{(\frac
{D-1}{2})}}\int_{\frac{m}{a}}^{\infty}dx{(\frac{a^{2}}{m^{2}}x^{2}-1)}%
^{\frac{D-3}{2}}K_{i2\omega/a}{(2x),} \label{mft}%
\end{equation}
and rewrite the Fourier transform of (\ref{moe}) as
\begin{equation}
{\mathcal{G}}^{(b)}=\frac{am^{D-4}}{L{(4\pi)}^{\frac{D-1}{2}}\Gamma
{(\frac{D-1}{2})}}\int_{\frac{m}{a}}^{\infty}dx{(\frac{a^{2}}{m^{2}}x^{2}%
-1)}^{\frac{D-4}{2}}\sin{[mL{(\frac{a^{2}}{m^{2}}x^{2}-1)}^{1/2}]}%
K_{i2\omega/a}{(2x),} \label{mfto}%
\end{equation}
where we have used the integration formulae
\begin{equation}
\int_{-\infty}^{\infty}e^{-i{(2\omega_{k})}\sinh{au/2}}e^{i\omega u}%
du=\frac{4}{a}e^{\frac{\pi\omega}{a}}K_{i2\omega/a}{(2\omega_{k}/a).}%
\end{equation}
Note that the signs ${\mathcal{G}}^{(a)}$ and ${\mathcal{G}}^{(b)}$ represents
the Fourier form of the diagonal and off-diagonal terms for the Wightman functions.

\begin{figure}[ptb]
\centering
\includegraphics[width=7in,height=4.5in]{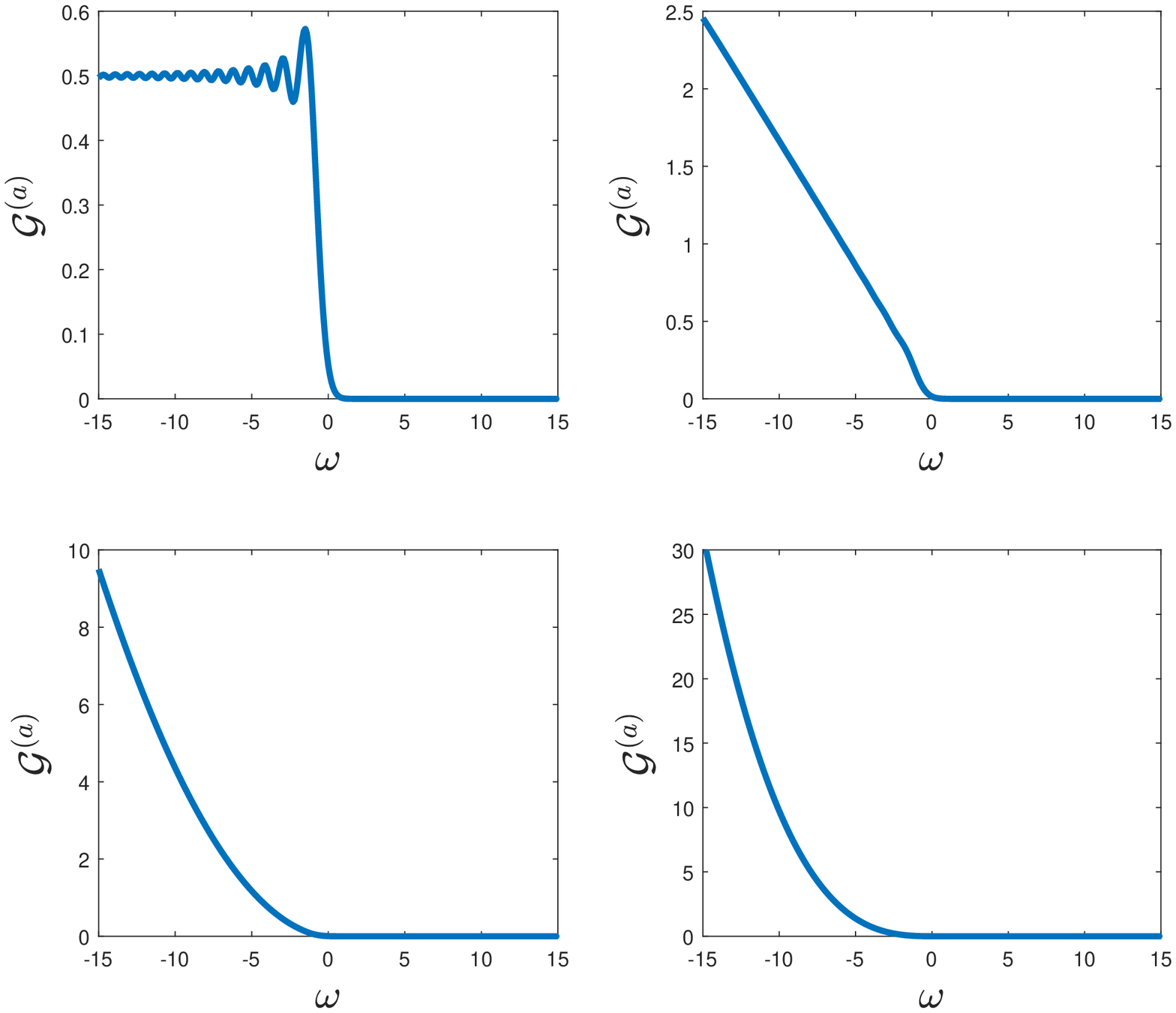}\caption{Behavior of
${\mathcal{G}}^{(a)}$ for massive field in different spacetime dimensions. The
four subfigures stand for $D = 3$, $D = 4$, $D = 5$, and $D = 6$,
respectively, from left to right and top to bottom.}%
\end{figure}

\begin{figure}[ptb]
\centering
\includegraphics[width=7in,height=4.5in]{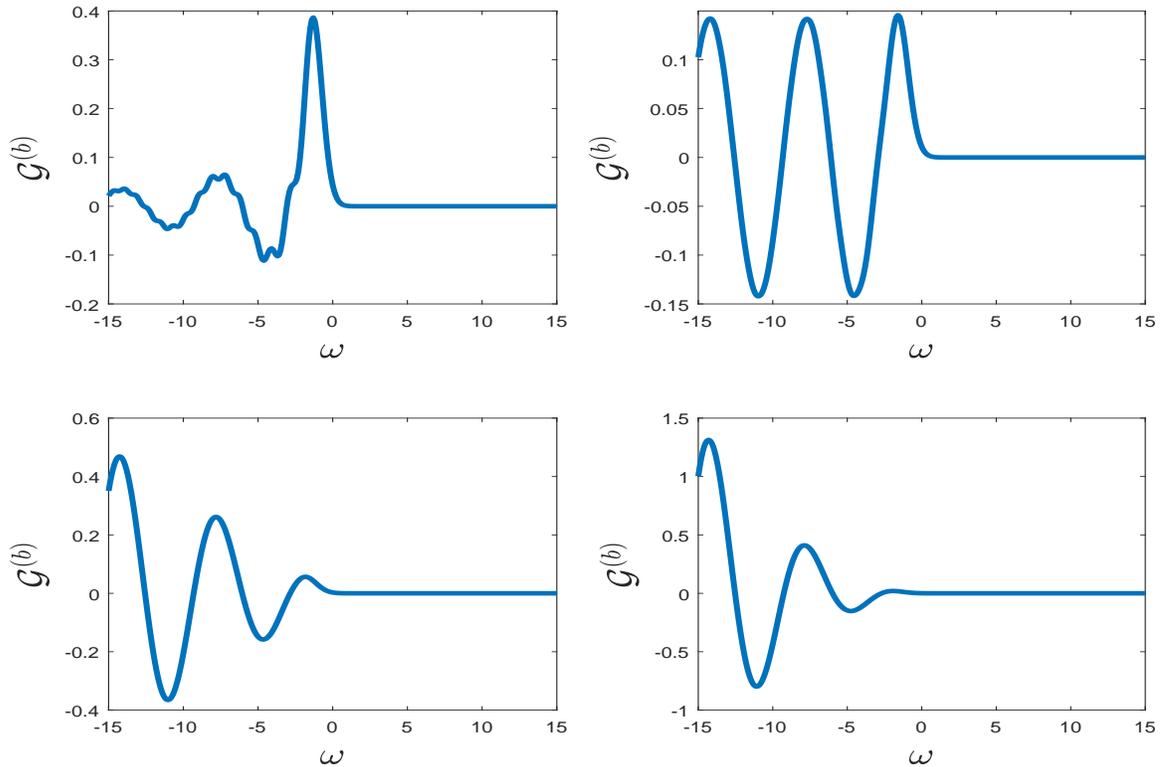}\caption{Behavior of
${\mathcal{G}}^{(b)}$ for massive field in different spacetime dimensions. The
four subfigures stand for $D = 3$, $D = 4$, $D = 5$, and $D = 6$,
respectively, from left to right and top to bottom.}%
\end{figure}

Fig. 11 and Fig. 12 show the power spectra in the case of massive field in
arbitrary spacetime dimensions by numerical integrating (\ref{mft}) and
(\ref{mfto}). Because of the modification of $K$-Bessel functions, all power
spectra for the diagonal terms decrease to zero gradually for any spacetime
dimension $D$, but it is special for the $D=3$ case that including the
oscillation in the axis of negative frequency and differs from that for other
spacetime dimensions, as given in Fig. 11. For the off-diagonal terms as given
in Fig. 12, the oscillation for the curves of the power spectra in different
spacetime dimensions has the same trend with the in Fig. 2 for the case of
massless field but the period of oscillation becomes larger for the case of
massive field due to the modulation of $K$-Bessel functions. Note that the
power spectra of off-diagonal terms for $D=3$ is also special, which, together
with the case of diagonal terms, derives from the function of statistical
effect, as discussed in the case of massless field. But for $D=4$, the
statistical effect is not evident for the case of massive field, which is
covered in the $K$-Bessel function. For $D>4$, the behaviors of the power
spectra for the cases of massless and massive field looks similar because the
statistical factor in these cases doesn't dominate.

\subsection{Entanglement change}

Adopting the method of the master equation described in the last section but
the power spectra are used with Eq. (\ref{mft}) and (\ref{mfto}), we calculate
entanglement dynamics of two accelerated atoms with the initial product state
for the case of the massive field under the different spacetime dimensions.
Fig. 13 presents the evolution of entanglement with time with the other
parameters taking $a,L,\omega,m=1$ for the cases of the massless and massive
field. By comparison, the entanglement evolution in the case of massive field
is always slower, which is an interesting time-delay phenomenon. It seems that
the speed of the massive field to propagate the interaction information
between two accelerated atoms is slower than that of massless field. Another
interesting phenomenon for the case of massive field is that the maximal
amount of generated entanglement decrease more quickly when the spacetime
dimension $D$ increases than that for the case of massless field. The cases
for $D=5$ and $D=6$ are consistent with these discussions here and don't
present any abnormal behaviors, so we don't plot them in Fig. 13.
\begin{figure}[ptb]
\centering
\includegraphics[width=5in,height=3in]{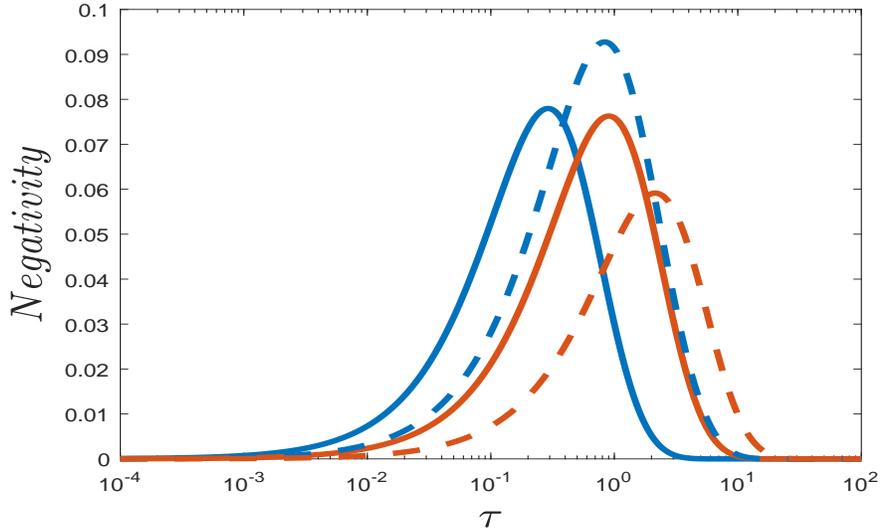}\caption{Entanglement
evolutionary process over time for massive field. Red and blue lines stand for
$D=3$ and $D=4$, respectively. Solid and dashed lines stand for the cases of
massless and massive field, respectively.}%
\end{figure}

Finally, let's discuss the anti-Unruh effect in different spacetime
dimensions. For this, we add an extra Gauss switching function $\chi
(\tau)=e^{-\tau^{2}/2\sigma^{2}}$ as made in Ref. \cite{bmm16,gmr16} and
calculate the Fourier transformation as
\begin{equation}
{\mathcal{G}}^{(a)}=\int d\tau d\tau^{\prime}G(\tau,\tau^{\prime}%
)e^{-i\omega(\tau-\tau^{\prime})}\chi(\tau)\chi(\tau^{\prime}).
\end{equation}
For simplicity, we take $\tau^{\prime}=0$ here, and obtain
\begin{equation}
{\mathcal{G}}^{(a)}=\int d\tau G(\tau)e^{-i\omega(\tau)}\chi(\tau).
\label{aufw}%
\end{equation}
This equation is proportional to the transition probability of an Unruh-Dewitt
detector \cite{bmm16,gmr16}. Inserting Eq. (\ref{mdem}) into Eq. (\ref{aufw}),
we can integrate and obtain the power spectrum by employing numerical method
(the oscillatory phenomenon we mentioned before does not matter now, because
the Gaussian function has suppressed the oscillation to a very low level). The
integral range was chosen as $[-10\sigma,10\sigma]$ ($\sigma=0.8$) to suppress
the numerical error at level of $e^{-50}\sim10^{-22}$ due to the finite
integration range. Fig. 14 presents the change of ${\mathcal{G}}^{(a)}$ with
acceleration and the anti-Unruh effect is feasible in any spacetime dimension.
Interestingly, the anti-Unruh effect is robust to the influence of the
spacetime dimensions. But for the Unruh effect in the $D=3$ and $D=4$ cases,
it changes into the anti-Unruh effect in the $D=5$ and $D=6$ cases with the
same parameters. This is a remarkable influence from the spacetime dimensions
but the fundamentally physical reason is unclear now and remains to be
explored in the future. In fact, the fundamentally physical reason for the
appearance of anti-Unruh effect is not clear up to now although the required
conditions has been discussed many times
\cite{bmm16,gmr16,lzy18,zhy21,chy22,pz20,pz21,bm21}.\begin{figure}[ptb]
\centering
\includegraphics[width=7in,height=4.5in]{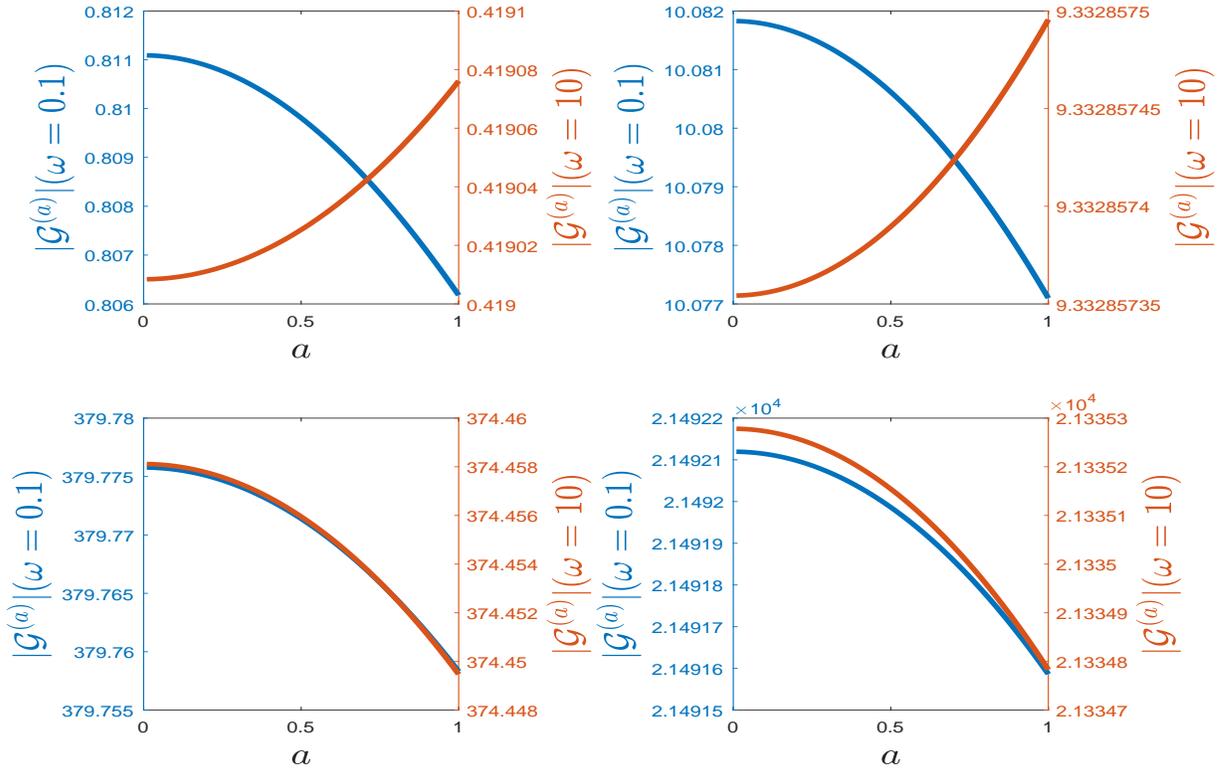}\caption{Anti-Unruh
effect in different spacetime dimensions. The four subfigures stand for $D=3$,
$D=4$, $D=5$, and $D=6$, respectively, from left to right and top to bottom.
Blue and red lines stand for $\omega=0.1$ and $\omega=10$, respectively, with
other parameters taking $m=0.1$, $\sigma=1$.}%
\end{figure}

\section{Conclusion}

In this paper, we have investigated the entanglement dynamics of a quantum
system composed of two uniformly accelerated Unruh-DeWitt detectors coupled
with both fluctuating massless and massive field in arbitrary dimensional
Minkowski vacuum. The main results focused on the influence of the spacetime
dimensions on the change of entanglement.

At first, we investigate the behaviors of the power spectrum for different
spacetime dimensions. For the case of the massless field, the power spectrum
obeys Bose-Einstein distribution in even dimensions, and Fermi-Dirac
distribution in odd dimensions. For diagonal components, the power spectrum
completely obeys Fermi-Dirac distribution when $D=3$, but has to be regarded
as the multiplication of a polynomial multiplication term and a Bose/Fermi
factor when $D\geq4$ in which the statistical effect doesn't dominate. For
off-diagonal components, the power spectrum can be regarded as the
multiplication of the corresponding power spectrum for diagonal components and
a hypergeometric function. The power spectra for off-diagonal components
present the oscillating behavior with $\omega$ but differs for different
specetime dimensions.

Then, the influence of spacetime dimensions on change of entanglement between
two accelerated atoms is studied in the case of the massless field.
Considering an initial product state for two atoms, the acceleration leads to
the generation of entanglement between two atoms. It is found that the range
of the parameters ($\omega,L,a$) in which entanglement between two atoms can
be generated is expanded with increasing spacetime dimension $D$, which is
measured by the area of entanglement region from Fig. 3 to Fig. 6. However,
the maximal amount of generated entanglement decreases with increasing $D$,
which is seen by the entanglement evolution with time in Fig. 7. We also study
the change of generated maximal entanglement with acceleration for different
atom separations and find that for the proper separation between two atoms,
entanglement can be enhanced for a certain range of increasing acceleration,
which is similar to the anti-Unruh effect, but this phenomenon doesn't appear
at the same separation for different spacetime dimensions and it appears at
smaller separation for lower spacetime dimensions. When $L$ is large enough,
the maximal entanglement will oscillate with increasing acceleration, which
shows that the generated entanglement is unstable and very small. When $L\ $is
increased further ($\rightarrow\infty$), the entanglement will decrease very
quickly to zeros in any spacetime dimensions. When the two atoms are placed at
the same places ($L\rightarrow0)$, entanglement is generated to the maximal
value at some certain $\omega$, remains there for a certain range of $\omega$,
and then decreases to zero suddenly at another certain $\omega$. Note that
these certain $\omega$ is dependent on the spacetime dimensions. Moreover, the
change of entanglement for the initial entangled state of two atom is also
studied, and the difference of entanglement change with time for different
spacetime dimensions are presented with different initial entangled states for
two atoms.

Finally, we expand the discussion to the case of the massive field. The
Wightman functions and the power spectra are obtained by adding an extra
factor expressed by K-Bessel functions in the basis of the results for the
case of the massless field. When entanglement evolution is calculated in this
case, it is found that mass can lead to a time delay for entanglement
generation in any spacetime dimensions, compared with that for the case of the
massless field. But the generated maximal entanglement decreases more quickly
with increasing spacetime dimension than that for the case of massless field.
When we study the change of generated maximal entanglement with the
acceleration, a surprised and attractive result is found that the Unruh effect
in the small spacetime dimensions can change to the anti-Unruh effect in large
spacetime dimensions with the same other parameters.

\acknowledgments

This work is supported by the NSFC Grant Nos. 11654001.


\begin{thebibliography}{99}                                                                                               %


\bibitem {wgu76}W. G. Unruh, \textit{Notes on black-hole evaporation,}
\textit{Phys. Rev. D} \textbf{14} (1976) 870.

\bibitem {pd75}P. C. W. Davies, \textit{Scalar production in Schwarzschild and
Rindler metrics,} \textit{J. Phys. A} \textbf{8} (1975) 609.

\bibitem {saf73}S. A. Fulling, \textit{Nonuniqueness of canonical field
quantization in Riemannian space-time.} \textit{Phys. Rev. D} \textbf{7}
(1973) 2850.

\bibitem {chm08}L. C. B. Crispino, A. Higuchi and G. E. A. Matsas, \textit{The
Unruh effect and its applications, Rev. Mod. Phys.} \textbf{80} (2008) 787.

\bibitem {cp11}M. Chernicoff and A. Paredes, \textit{Accelerated detectors and
worldsheet horizons in AdS/CFT,} \textit{JHEP} \textbf{1103} (2011) 063.

\bibitem {rt08}J. G. Russo and P. K. Townsend, \textit{Accelerating branes and
brane temperature,} \textit{Class. Quantum Grav.} \textbf{25} (2008) 175017.

\bibitem {aas06}A. A. Saharian, \textit{Wightman function and vacuum
fluctuations in higher dimensional brane models, Phys. Rev. D} \textbf{73}
(2006) 044012.

\bibitem {am95}J. Audretsch and R. M\"{u}ller, \textit{Radiative energy shifts
of an accelerated two-level system,} \textit{Phys. Rev. A} \textbf{52} (1995) 629.

\bibitem {rp98}R. Passante, \textit{Radiative level shifts of an accelerated
hydrogen atom and the Unruh effect in quantum electrodynamics,} \textit{Phys.
Rev. A} \textbf{57} (1998) 1590.

\bibitem {rzp16}L. Rizzuto, M. Lattuca, J. Marino, A. Noto, S. Spagnolo, W.
Zhou and R. Passante, \textit{Nonthermal effects of acceleration in the
resonance interaction between two uniformly accelerated atoms,} \textit{Phys.
Rev. A} \textbf{94} (2016) 012121.

\bibitem {zpr16}W. Zhou, R. Passante and L. Rizzuto, \textit{Resonance
interaction energy between two accelerated identical atoms in a coaccelerated
frame and the Unruh effect,} \textit{Phys. Rev. D} \textbf{94} (2016) 105025.

\bibitem {fm05}I. Fuentes-Schuller and R. B. Mann, \textit{Alice Falls into a
Black Hole: Entanglement in Noninertial Frames, Phys. Rev. Lett.} \textbf{95}
(2005) 120404.

\bibitem {dss15}Y. Dai, Z. Shen, and Y. Shi, \textit{Killing quantum
entanglement by acceleration or a black hole, JHEP} \textbf{09} (2015) 071.

\bibitem {amt06}P. M. Alsing, I. Fuentes-Schuller, R. B. Mann and T. E.
Tessier, \textit{Entanglement of Dirac fields in noninertial frames,}
\textit{Phys. Rev. A} \textbf{74} (2006) 032326.

\bibitem {ml09}E. Mart\'{\i}n-Mart\'{\i}nez and J. Le\'{o}n, \textit{Fermionic
entanglement that survives a black hole}, \textit{Phys. Rev. A} \textbf{80}
(2009) 042318.

\bibitem {mgl10}E. Mart\'{\i}n-Mart\'{\i}nez, L. J. Garay and J. Le\'{o}n,
\textit{Unveiling quantum entanglement degradation near a Schwarzschild black
hole,} \textit{Phys. Rev. D} \textbf{82} (2010) 064006.

\bibitem {wj11}J. Wang and J. Jing, \textit{Multipartite entanglement of
fermionic systems in noninertial frames,} \textit{Phys. Rev. A} \textbf{83}
(2011) 022314.

\bibitem {ses12}M. Shamirzaie, B. N. Esfahani and M. Soltani,
\textit{Tripartite entanglements in noninertial frames, Int. J. Theor. Phys.}
\textbf{51} (2012) 787.

\bibitem {bfl12}D. E. Bruschi, A. Dragan, I. Fuentes and J. Louko,
\textit{Particle and antiparticle bosonic entanglement in noninertial frames,
Phys. Rev. D} \textbf{86} (2012) 025026.

\bibitem {ro15}B. Richter and Y. Omar, \textit{Degradation of entanglement
between two accelerated parties: Bell states under the Unruh effect, Phys.
Rev. A} \textbf{92} (2015) 022334.

\bibitem {hy15}J. Hu, and H. Yu, \textit{Entanglement dynamics for uniformly
accelerated two-level atoms, Phys. Rev. A} \textbf{91} (2015) 012327.

\bibitem {rrs05}B. Reznik, A. Retzker and J. Silman, \textit{Violating Bell's
inequalities in vacuum, Phys. Rev. A} \textbf{71} (2005) 042104.

\bibitem {sm09}G. V. Steeg and N. C. Menicucci, \textit{Entangling power of an
expanding universe, Phys. Rev. D} \textbf{79} (2009) 044027.

\bibitem {mm11}M. Montero and E. Mart\'{\i}n-Mart\'{\i}nez, \textit{The
entangling side of the Unruh-Hawking effect, JHEP} \textbf{07} (2011) 006.

\bibitem {dhi79}B. S. DeWitt, S. Hawking and W. Israel, \textit{General
relativity: an Einstein centenary survey,} Cambridge Press, Cambridge U. K. (1979).

\bibitem {br03}B. Reznik, \textit{Entanglement from the vacuum, Foundations of
Physics} \textbf{33} (2003) 167.

\bibitem {ctm19}W. Cong, E. Tjoa, R. B. Mann, \textit{Entanglement harvesting
with moving mirrors, JHEP} \textbf{06} (2019) 021.

\bibitem {zy20}J. Zhang, H. Yu, En\textit{tanglement harvesting for
Unruh-DeWitt detectors in circular motion, Phys. Rev. D} \textbf{102} (2020) 065013.

\bibitem {gtm21}K. Gallock-Yoshimura, E. Tjoa, R. B. Mann, \textit{Harvesting
entanglement with detectors freely falling into a black hole, Phys. Rev. D}
104 (2021) 025001.

\bibitem {cm22}P. Chowdhury and B. R. Majhiy, \textit{Fate of entanglement
between two Unruh-DeWitt detectors due to their motion and background
temperature, JHEP} 05 (2022) 025.

\bibitem {bmm16}W. G. Brenna, R. B. Mann and E. Mart\'{\i}n-Mart\'{\i}nez,
\textit{Anti-Unruh phenomena, Phys. Lett. B} \textbf{757} (2016) 307.

\bibitem {gmr16}L. J. Garay, E. Mart\'{\i}n-Mart\'{\i}nez and J. de Ram\'{o}n,
\textit{Thermalization of particle detectors: The Unruh effect and its
reverse,} \textit{Phys. Rev. D} \textbf{94} (2016) 104048.

\bibitem {rk57}R. Kubo, \textit{Statistical-mechanical theory of irreversible
processes. I. General theory and simple applications to magnetic and
conduction problems, J. Phys. Soc. Jpn.} \textbf{12} (1957) 570.

\bibitem {ms59}P. C. Martin and J. Schwinger, \textit{Theory of many-particle
systems. I, Phys. Rev.} \textbf{115} (1959) 1342.

\bibitem {fjl16}C. J. Fewster, B. A. Ju\'{a}rez-Aubry, and J. Louko,
\textit{Waiting for Unruh, Class. Quantum Grav.} \textbf{33} (2016) 165003.

\bibitem {lzy18}T. Li, B. Zhang and L. You, \textit{Would quantum entanglement
be increased by anti-Unruh effect? Phys. Rev. D} \textbf{97} (2018) 045005.

\bibitem {zhy21}Y. Zhou, J. Hu and H. Yu, \textit{Entanglement dynamics for
Unruh-DeWitt detectors interacting with massive scalar fields: the Unruh and
anti-Unruh effects, JHEP} \textbf{09} (2021) 088.

\bibitem {chy22}Y. Chen, J. Hu and H. Yu, \textit{Entanglement generation for
uniformly accelerated atoms assisted by environment-induced interatomic
interaction and the loss of the anti-Unruh effect, Phys. Rev. D} \textbf{105}
(2022) 045013.

\bibitem {pz20}Y. Pan and B. Zhang, \textit{Influence of acceleration on
multibody entangled quantum states, Phys. Rev. A} \textbf{101} (2020) 062111.

\bibitem {pz21}Y. Pan and B. Zhang. \textit{Anti-Unruh effect in the thermal
background,} \textit{Phys. Rev. D} \textbf{104} (2021) 125014.

\bibitem {bm21}S. Barman and B. R. Majhi, \textit{Radiative process of two
entangled uniformly accelerated atoms in a thermal bath: a possible case of
anti-Unruh event, JHEP} \textbf{03} (2021) 245.

\bibitem {btz92}M. Ba\~{n}nados, C. Teitelboim and J. Zanelli, \textit{The
Black Hole in Three Dimensional Spacetime, Phys. Rev. Lett.} \textbf{69}
(1992) 1849.

\bibitem {hhz20}L. J. Henderson, R. A. Hennigar, R. B. Mann, A. R. H. Smith,
J. Zhang, \textit{Anti-Hawking phenomena, Phys. Lett. B} \textbf{809} (2020) 135732.

\bibitem {rhm21}M. P. G. Robbins, L. J. Henderson and R. B. Mann,
\textit{Entanglement amplification from rotating black holes, Class. Quantum
Grav.} \textbf{39} (2021) 02LT01.

\bibitem {dss16}Y. Dai, Z. Shen and Y. Shi, \textit{Quantum entanglement in
three accelerating qubits coupled to scalar fields,} \textit{Phys. Rev. D}
\textbf{94} (2016) 025012.

\bibitem {wcj20}J. Wang, L. Zhang, S. Chen and J. Jing, \textit{Estimating the
Unruh effect via entangled many-body probes, Phys. Lett. B} \textbf{802}
(2020) 135239.

\bibitem {st86}S. Takagi, \textit{Vacuum noise and stress induced by uniform
accelerationhawking-unruh effect in rindler manifold of arbitrary dimension,
Prog. Theor. Phys. Suppl.} \textbf{88} (1986) 1.

\bibitem {ls02}L. Sriramkumar, \textit{Odd statistics in odd dimensions for
odd couplings, Mod. Phys. Lett. A} \textbf{17} (2002) 1059.

\bibitem {so17}S. Ohya, \textit{Emergent anyon distribution in the Unruh
effect, Phys. Rev. D} \textbf{96} (2017) 045017.

\bibitem {hy12}J. Hu and H. Yu, \textit{Geometric phase for an accelerated
two-level atom and the Unruh effect, Phys. Rev. A} \textbf{85} (2012) 032105.

\bibitem {fzz221}J. Feng, J. Zhang and Q. Zhang, \textit{Geometric phase under
the Unruh effect with intermediate statistics, Chin. Phys. B} \textbf{31}
(2022) 050312.

\bibitem {fz22}J. Feng and J. Zhang, \textit{Quantum Fisher information as a
probe for Unruh thermality, Phys. Lett. B} \textbf{827} (2022) 136992.

\bibitem {rlc17}J. Rodr\'{\i}uez-Laguna, L. Tarruell, M. Lewenstein and A.
Celi, \textit{Synthetic Unruh effect in cold atoms, Phys. Rev. A} \textbf{95}
(2017) 013627.

\bibitem {klc18}A. Kosior, M. Lewenstein and A. Celi, \textit{Unruh effect for
interacting particles with ultracold atoms, SciPost Phys.} \textbf{5} (2018) 061.

\bibitem {bp02}H.-P. Breuer and F. Petruccione, \textit{The theory of open
quantum systems,} Oxford University Press, Oxford U. K. (2002).

\bibitem {kb20}G. Kaplanek, and C. P. Burgess, \textit{Hot accelerated qubits:
decoherence, thermalization, secular growth and reliable late-time
predictions, JHEP} \textbf{03} (2020) 008.

\bibitem {abg21}Julio Arrechea, Carlos Barcel\'{o}, L. J. Garay and G.
Garc\'{\i}a-Moreno, \textit{Inversion of statistics and thermalization in the
Unruh effect, Phys. Rev. D} \textbf{104} (2021) 065004.

\bibitem {jz07}A. Jeffrey and D. Zwillinger, eds, \textit{Table of integrals,
series, and products,} Academic Press, Elsevier, Burlington U. S. A. (2007).

\bibitem {rhd54}R. H. Dicke, \textit{Coherence in Spontaneous Radiation
Processes, Phys. Rev.} 93 (1954) 99.

\bibitem {vw02}G. Vidal, R. F. Werner, \textit{Computable measure of
entanglement, Phys. Rev. A} \textbf{65} (2002) 032314.

\bibitem {zsl98}K. Zyczkowski, P. Horodecki, A. Sanpera, M. Lewenstein,
\textit{On the volume of the set of mixed entangled states}, \textit{Phys.
Rev. A} \textbf{58} (1998) 883.

\bibitem {ap96}A. Peres, \textit{Separability criterion for density matrices,
Phys. Rev. Lett.} \textbf{77} (1996) 1413.

\bibitem {hhh01}M. Horodecki, P. Horodecki and R. Horodecki,
\textit{Separability of n-particle mixed states: necessary and sufficient
conditions in terms of linear maps, Phys. Lett. A} \textbf{283} (2001) 1.
\end{thebibliography}
\end{document}